\documentclass[twocolumn,showkeys,preprintnumbers,amsmath,amssymb,10pt]{revtex4-1}

\usepackage{geometry}
\usepackage{graphicx}
\usepackage{amssymb}
\usepackage{amsmath}
\usepackage{upgreek}
\usepackage{multirow}
\usepackage{dcolumn}
\usepackage{bm}
\usepackage[latin1]{inputenc}
\usepackage{color}
\usepackage[
 breaklinks=true
 ,colorlinks=true
 ,linkcolor=blue 
 ,citecolor=blue
 ,backref=none
 ,pagebackref=false
 ]{hyperref} 

\usepackage{setspace}
\usepackage{epstopdf}
\pdfoutput=1
\usepackage{geometry}
\usepackage{setspace} 
\usepackage[mathlines]{lineno}
\usepackage{todonotes}

\makeatletter
\renewcommand{\@biblabel}[1]{\quad#1.}
\makeatother

\newcommand{\olN}{\overline N}
\newcommand{\olpi}{\overline \pi}
\newcommand{\oli}{\overline i}
\newcommand{\olj}{\overline j}

\newcommand{\avg}[1]{\left\langle{#1}\right\rangle}
\renewcommand{\d}{{\rm d}}

\begin{document}


\title{Effects of population growth on the success of invading mutants}
\author{Peter Ashcroft}\thanks{peter.ashcroft@env.ethz.ch}
\affiliation{ETH Z\"urich, Institut f\"ur Integrative Biologie, 8092 Z\"urich, Switzerland}

\author{Cassandra E. R. Smith}\thanks{Current address: Institute of Neuroscience, Newcastle University, Newcastle upon Tyne, NE2 4HH, UK, c.smith26@newcastle.ac.uk.}
\author{Matthew Garrod}\thanks{Current address:  Department of Mathematics, Imperial College London, London
W7 2AZ, UK, m.garrod15@imperial.ac.uk.}

\author{Tobias Galla}\thanks{tobias.galla@manchester.ac.uk}

\affiliation{Theoretical Physics, School of Physics and Astronomy, The University of Manchester, Manchester M13 9PL, UK}


\label{firstpage} 




\begin{abstract}
Understanding if and how mutants reach fixation in populations is an important question in evolutionary biology. We study the impact of population growth has on the success of mutants. To systematically understand the effects of growth we decouple competition from reproduction; competition follows a birth--death process and is governed by an evolutionary game, while growth is determined by an externally controlled branching rate. In stochastic simulations we find non-monotonic behaviour of the fixation probability of mutants as the speed of growth is varied; the right amount of growth can lead to a higher success rate.
These results are observed in both coordination and coexistence game scenarios, and we find that the `one-third law' for coordination games can break down in the presence of growth.
We also propose a simplified description in terms of stochastic differential equations to approximate the individual-based model.
\end{abstract}
 
\maketitle



\section{Introduction}
When and how mutants spread in wildtype populations is an important question in population dynamics; answering it has implications in bacterial evolution, cancer initiation, viral dynamics and for the understanding of social phenomena \cite{maddamsetti:Genetics:2015,nowak:book:2006,altrock:NatRev:2015,castellano:RMP:2009}.
While the behaviour of populations has traditionally been described mostly with deterministic models \cite{hofbauer:book:1998,maynard-smith:book:1982}, it is increasingly recognised that the fate of invading mutants can be influenced by random genetic drift.
Work from recent decades reflects this shift in modelling, and much current research is concerned with the properties of stochastic evolution in finite populations \cite{nowak:book:2006,traulsen:bookchapter:2009,ewens:book:2004,goel:book:1974,taylor:BMB:2004,bladon:PRE:2010}.
 
Mathematical models of stochastic evolution typically describe a population of individuals who can each be of one of several types or species.
In the simplest scenario one considers the spread of mutant individuals in a wildtype population.
Often the interactions between species follow a birth--death process; individuals of one type may generate offspring at the expense of other individuals who are removed from the population, such that the total population size is conserved.
These events occur stochastically and their rates are determined by the relative reproductive fitnesses of the different species; these fitnesses in turn depend on the composition of the population \cite{nowak:book:2006,traulsen:bookchapter:2009}.
Evolutionary game theory is a commonly-used framework for describing these frequency-dependent dynamics.
Fixation probabilities and mean fixation times can be computed for these stylised models using techniques from the theory of stochastic processes \cite{gardiner:book:2009,kampen:book:2007,ewens:book:2004,altrock:NJP:2009,antal:BMB:2006}.

More recently, work has also focused on models with populations of dynamic size.
Melbinger \emph{et al.} have investigated the impact that demography has on the spread of cooperation in the prisoners dilemma game \cite{melbinger:PRL:2010,cremer:PRE:2011,cremer:SciRep:2012}.
Other evolutionary game formats have also been studied in populations of time-dependent size \cite{novak:JTB:2013,chotibut:PRE:2015,huang:PNAS:2015,li:Interface:2015,constable:PNAS:2016}.
In these models growth is limited by an overall carrying capacity.
The effect of population growth has also been considered in host--parasite interactions \cite{papkou:Zoo:2016}, and in spin systems \cite{morris:JPhysA:2014}.
Other work specifically focuses on range expansions in space \cite{hallatschek:PNAS:2007}.

\begin{figure}
\centering
\includegraphics[width=\columnwidth]{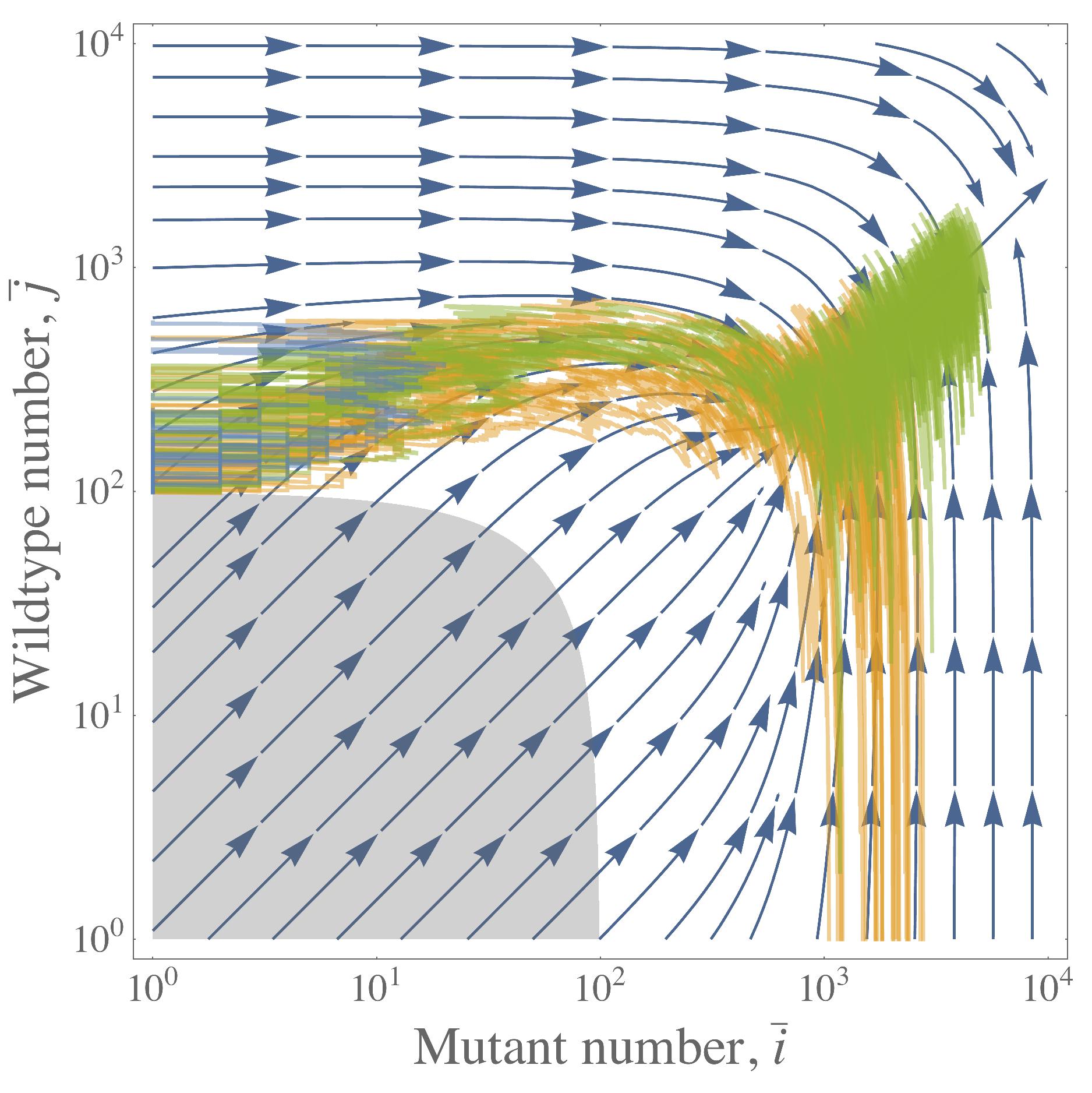}
\caption{
Stochastic trajectories of a single mutant in a growing population subject to coexistence-game dynamics.
Trajectory colours indicate the outcome of extinction, fixation, or indefinite coexistence.
Background arrows are the deterministic flow field, as described in section~\ref{sec:det}.
Data shown is for selection strength $\beta = 0.1$, payoff matrix~\eqref{eq:payoffCoex} with fixed point $x^\star=0.8$, initial population size $N_0=100$, and power-law population growth with exponent $\gamma=0.4$.
}
\label{fig:0-trajectories}
\end{figure}

In such models that combine selection and growth, an interesting interplay between the underlying deterministic dynamics and intrinsic noise is to be expected.
For example, consider a scenario where the deterministic flow has a stable fixed point for non-zero numbers of both types of individual.
For infinite populations noise can be neglected, and the deterministic flow leads to the indefinite coexistence of two species.
In finite populations, however, extinction of one of the two types can and will occur as these phenomena are driven by the intrinsic noise.
A growing population presents an interesting intermediate case; if its initial size is small, demographic stochasticity shapes the outcome in the early phases (in populations of size $N$, intrinsic noise has an amplitude of order $N^{-1/2}$ relative to deterministic selection).
As the population grows the relative strength of the noise gradually reduces, and in the latter stages deterministic flow dominates over random drift.
This can lead to outcomes of fixation or extinction, or indefinite coexistence, as highlighted in Fig.~\ref{fig:0-trajectories}.
The speed of growth determines how long intrinsic noise is relevant before the deterministic flow takes over.
The purpose of our work is to investigate this in more detail and to characterise the outcome of evolution for different speeds of population growth.

To address this issue we explicitly decouple the between-species interactions -- birth--death dynamics in the form of a two-player two-strategy evolutionary game -- from the reproduction dynamics leading to population growth.
We consider evolutionary scenarios described by the well-known cases of the dominance, coordination, and coexistence games \cite{traulsen:bookchapter:2009}, as described in Sec.~\ref{sec:model:competition}.
Growth in our model is governed by an externally-controlled per-capita reproduction rate, $\Gamma(N,t)$, which may depend on the current population size and/or have an explicit time dependence as described in Sec.~\ref{sec:model:growth}.
This rate is not frequency-dependent, such that the growth process itself does not favour any of the two species -- selection is controlled only by the between-species interactions.
By varying the growth law independently from the selection dynamics, we can systematically test the effect of population growth on the evolutionary outcome.
This is much harder to do in models in which growth and selection are combined, as the growth law then `emerges' from the population itself and cannot easily be controlled externally.

\section{Model definitions} \label{sec:model}

We consider a well-mixed, growing population of discrete individuals.
Each member of the population can be one of two types, $A$ or $B$.
We will refer to species $A$ as the `mutant' type, and $B$ as the `wildtype'.
The state of the population at any time can be described by the pair of variables $(i,N)$.
In our notation $i(t)$ is the number of individuals of type $A$ (mutants), and $N(t)$ is the total number of individuals in the population at a given time $t$.
The number of individuals of type $B$ (wildtypes) can be written as $j(t) = N(t) - i(t)$.
Generally we are interested in the fate of a single mutant in a population of wildtype individuals.
For the dynamics we consider two types of discrete events: Competition between the species, and growth of the population.

\subsection{Competition} \label{sec:model:competition}
Competition (or selection) in our model is governed by transitions of the type $(i,N) \to (i \pm 1,N)$, i.e. replacement of an individual by another of the opposite type without increasing the population size through a birth--death process.
We use the framework of evolutionary game theory to describe these interactions \cite{nowak:book:2006,traulsen:bookchapter:2009}.
When two individuals interact, the likelihood for one type to succeed over the other is characterised by the individuals' expected payoffs within the population.
These are written as $\pi_A(i,N)$ and $\pi_B(i,N)$ respectively for members of the two types; their precise form will be defined below.
In our model the rates at which these selection events happen are given by
\begin{linenomath}
\begin{align}
(i,N) \to (i + 1,N): \quad T_{i,N}^+ &= \frac{i(N-i)}{N} g(\pi_A,\pi_B) \nonumber \\
(i,N) \to (i - 1,N): \quad T_{i,N}^- &= \frac{i(N-i)}{N} g(\pi_B,\pi_A).
\label{eq:bdRates}
\end{align}
\end{linenomath}
This follows the lines, for example, of \cite{bladon:PRE:2010,traulsen:bookchapter:2009}.
The detailed mechanics of these birth--death events are governed by the competition kernel $g(\cdot,\cdot)$.
Generally, this kernel will be an increasing function of the first argument, and decreasing in the second.
For our investigation we use the common choice
\begin{equation}
g(\pi_A,\pi_B) = \frac{1}{1+\exp[-\beta(\pi_A-\pi_B)]},
\end{equation}
which is sometimes referred to as the `Fermi process' \cite{traulsen:bookchapter:2009,bladon:PRE:2010,altrock:NJP:2009}.
The parameter $\beta \geq 0$ is the intensity of selection: For $\beta=0$ evolution is neutral with no selection bias in favour of either species, for non-zero values of $\beta$ the payoffs determine the direction of selection.

We focus on the case of frequency-dependent selection; the expected payoffs of the two species depend on the current composition of the population.
As is often done in the existing literature (see e.g. \cite{nowak:book:2006}) we assume that $\pi_A(i,N)$ and $\pi_B(i,N)$ are determined by the following payoff matrix and functions:
\begin{equation}
\begin{array}{c|c c} & A & B \\\hline A & a & b \\ B & c & d \end{array},
\quad
\begin{array}{l}
\pi_A(i,N) = \dfrac{i-1}{N-1}a + \dfrac{N-i}{N-1}b, \\[1em]
\pi_B(i,N) = \dfrac{i}{N-1}c + \dfrac{N-i-1}{N-1}d.
\end{array}
\end{equation}
The parameter $a$ describes the payoff an individual of type $A$ receives from an interaction with another individual of type $A$.
Parameter $b$ is the payoff to $A$ when interacting with an individual of type $B$.
Parameters $c$ and $d$ follow similarly.

For these interactions alone (i.e. without changes in population size), the deterministic dynamics can be described by the so-called replicator equation \cite{hofbauer:book:1998}.
Writing $x(t) = \avg{i(t)/N(t)} = \avg{i(t)}/N$, where $\avg{\cdot}$ represents an average over infinitely many realisations of the stochastic process, and assuming that higher-order moments factorise (e.g. $\avg{i^2} = \avg{i}^2$), we have
\begin{equation}
\dot{x} = x(1-x)\left[g(\olpi_A,\olpi_B)-g(\olpi_B,\olpi_A)\right].
\label{eq:replicator}
\end{equation}
Here $g(\olpi_A,\olpi_B)$ [$g(\olpi_B,\olpi_A)$] is the (effective) fitness of type $A$ [$B$], where $\olpi_A(x) = x a + (1-x)b$ and $\olpi_B(x) = x c + (1-x)d$.
The choice of the payoff matrix elements $a$, $b$, $c$, and $d$ determines the shape of the selection bias.
Our analysis below focuses on several types of games, representing the different structures that can arise.
These include cases in which one species strictly dominates the other (Eq.~\eqref{eq:replicator} has no fixed points for $0<x<1$), as well as cases that promote coexistence of the species (stable internal fixed point) or lead to bistability (unstable internal fixed point) in coordination games  \cite{hofbauer:book:1998}.

\subsection{Growth} \label{sec:model:growth}
Population growth occurs through transitions $(i,N) \to (i+1,N+1)$ and $(i,N) \to (i,N+1)$.
In the former case an individual of type $A$ (mutant) reproduces, in the latter an individual of type $B$ (wildtype) generates an offspring.
These processes occur without the removal of any existing member of the population, such that the overall size of the population increases.
They happen with rates
\begin{linenomath}
\begin{align}
(i,N) \to (i+1,N+1): \quad r_{i,N}^A &= i \Gamma(N,t),\nonumber \\
(i,N) \to (i,N+1): \quad r_{i,N}^B &= (N-i) \Gamma(N,t),
\label{eq:growthevents}
\end{align}
\end{linenomath}
respectively.
The per-capita growth rate, $\Gamma(N,t)$, is assumed to be the same for both species such that there is no selection in this sector of the model.

We first consider the deterministic behaviour of the population size.
Writing $\olN(t) = \avg{N(t)}$, we find
\begin{equation}
\frac{\d \olN}{\d t} = \olN \, \Gamma(\olN,t).
\label{eq:detN}
\end{equation}
Therefore, we can use $\Gamma(N,t)$ as a mathematical device to generate specific growth laws for the average population size.
For example, choosing $\Gamma(N,t) = \alpha$ (const.) corresponds to exponential growth, or $\Gamma(N,t) = r(1-N/K)$ generates logistic growth with carrying capacity $K$.

For the rest of the manuscript we study a population whose size follows a power-law in time,
\begin{equation}
\olN(t) = N_0 t^\gamma.
\label{eq:powerLaw}
\end{equation}
This is achieved by setting
\begin{equation}
\Gamma(N,t) = \frac{\gamma}{t},
\label{eq:reproductionRate}
\end{equation}
or likewise
\begin{equation}
\Gamma(N,t) = \gamma\left(\frac{N_0}{N}\right)^{1/\gamma}.
\label{eq:reproductionRateN}
\end{equation}
We always assume that the dynamics is started at time $t=1$ with an initial size $N(t=1)=N_0$, and that the growth exponent $\gamma \geq 0$.
This growth law captures a variety of behaviours: Choosing $\gamma = 0$ corresponds to a scenario with a constant population size, $N(t) = N_0$.
On the other hand, choosing $\gamma = 1$ results in linear growth over time.
In the region $0 < \gamma < 1$ (sub-linear), the population initially experiences rapid growth, as is common when population densities are low.
As the size increases, the growth becomes suppressed (although the population increases indefinitely).
Our results below show that the choice of power-law growth captures non-trivial behaviour; in particular we find that a single mutant can be most successful at intermediate choices of the growth exponent.

Power-law growth has been observed in tumour development \cite{altrock:NatRev:2015}, where surface area \cite{cappuccio:CancerRes:2006} or radii \cite{bru:BPJ:2003} grow linearly in time.
Recently, Karev has highlighted the ubiquity of power-law growth across many scales of natural processes \cite{karev:MathBiosci:2014}, including the sub-exponential growth of replicators (nucleotides) \cite{szathmary:JTB:1997}.

This choice of growth law can also be motivated physically as follows:
Under simple birth--death dynamics the variance of the number of mutants grows linearly in time \cite{redner:book:2001}.
We can expect that the number of mutants in different realisations of the process will differ within a range proportional to the standard deviation, growing as $t^{1/2}$.
Imposing additional population growth, with no advantage to either species, effectively induces a growing boundary for this random-walk problem.
The most interesting scenarios are to be expected when the growth of the domain (i.e. the population size) and the typical deviations from the mean of mutant numbers follow similar laws.
The boundaries of the domain are then `reachable', so that fixation and extinction may occur.
On the other hand, intrinsic noise does not dominate when the population is large so that arrival at either of the boundaries may not necessarily be certain.

Finally, although the two representations of the per-capita growth rate [Eqs.~\eqref{eq:reproductionRate} and \eqref{eq:reproductionRateN}] result in the same dynamics, their interpretations are different.
The explicit time-dependence of Eq.~\eqref{eq:reproductionRate} suggests external moderation of the growth rate, such as controlling nutrient supply in an experiment.
On the other hand, the appearance of the total population size in Eq.~\eqref{eq:reproductionRateN} suggests some degree of self-moderation, similar to logistic or Gompertzian growth.


\section{Deterministic flow for $2 \times 2$ games in growing populations} \label{sec:det}
The full model is stochastic and individual-based.
A mathematical description can be formed in terms of the master equation which describes the behaviour of the probability, $P_{i,N}(t)$, to find the population in state $(i,N)$ at time $t$.
It is given by
\begin{linenomath}
\begin{align}
\frac{\d P_{i,N} }{\d t} 
=&~T_{i-1,N}^+ P_{i-1,N} - T_{i,N}^+ P_{i,N}\nonumber \\
& + T_{i+1,N}^- P_{i+1,N} - T_{i,N}^- P_{i,N}\nonumber \\
& + r_{i-1,N-1}^A P_{i-1,N-1} - r_{i,N}^A P_{i,N} \nonumber \\
& + r_{i,N-1}^B P_{i,N-1} - r_{i,N}^B P_{i,N}.
\end{align}
\end{linenomath}
From this equation we can derive ordinary differential equations (ODEs) for the time-evolution of the first moments, $\oli = \avg{i(t)}$, $\olN = \avg{N(t)}$.
This follows standard steps as described, for example, in \cite{kampen:book:2007,gardiner:book:2009}.
The population size, $\olN$, follows Eq.~\eqref{eq:detN}.
The number of mutants, $\oli$, satisfies
\begin{equation}
\frac{\d \oli}{\d t} = \frac{\oli(\olN-\oli)}{\olN} \left[g(\olpi_A,\olpi_B) - g(\olpi_B,\olpi_A)\right] + \oli\, \Gamma(\olN, t).
\label{eq:deti}
\end{equation}
We can show the fraction of mutants, $x=\oli/\olN$, follows the replicator equation~\eqref{eq:replicator} by using Eqs.~\eqref{eq:detN}, \eqref{eq:deti}, and the quotient rule of differentiation: $\dot{x} = \dot{\oli}/\olN - x \dot{\olN}/\olN$.

We can represent the deterministic dynamics in the ($\oli$,$\olj$)-plane, where $\olj = \olN - \oli$ is the number of wild-type individuals.
We use Eq.~\eqref{eq:reproductionRateN} to express $\Gamma$ in terms of $\olN = \oli + \olj$, and we arrive at closed expressions for the evolution of $\oli$ and $\olj$,
\begin{linenomath}
\begin{align}
\frac{\d \oli}{\d t} &= \frac{\oli \, \olj}{\oli + \olj} \left[g(\olpi_A,\olpi_B) - g(\olpi_B,\olpi_A)\right] + \gamma \oli \left(\frac{N_0}{\oli + \olj}\right)^{1/\gamma}, \nonumber\\
\frac{\d \olj}{\d t} &= \frac{\oli \, \olj}{\oli + \olj} \left[g(\olpi_B,\olpi_A) - g(\olpi_A,\olpi_B)\right] + \gamma \olj \left(\frac{N_0}{\oli + \olj}\right)^{1/\gamma}.
\label{eq:detijPower}
\end{align}
\end{linenomath}
We now consider these dynamics under different evolutionary game scenarios.

\begin{figure}
\centering
\includegraphics[width=\columnwidth]{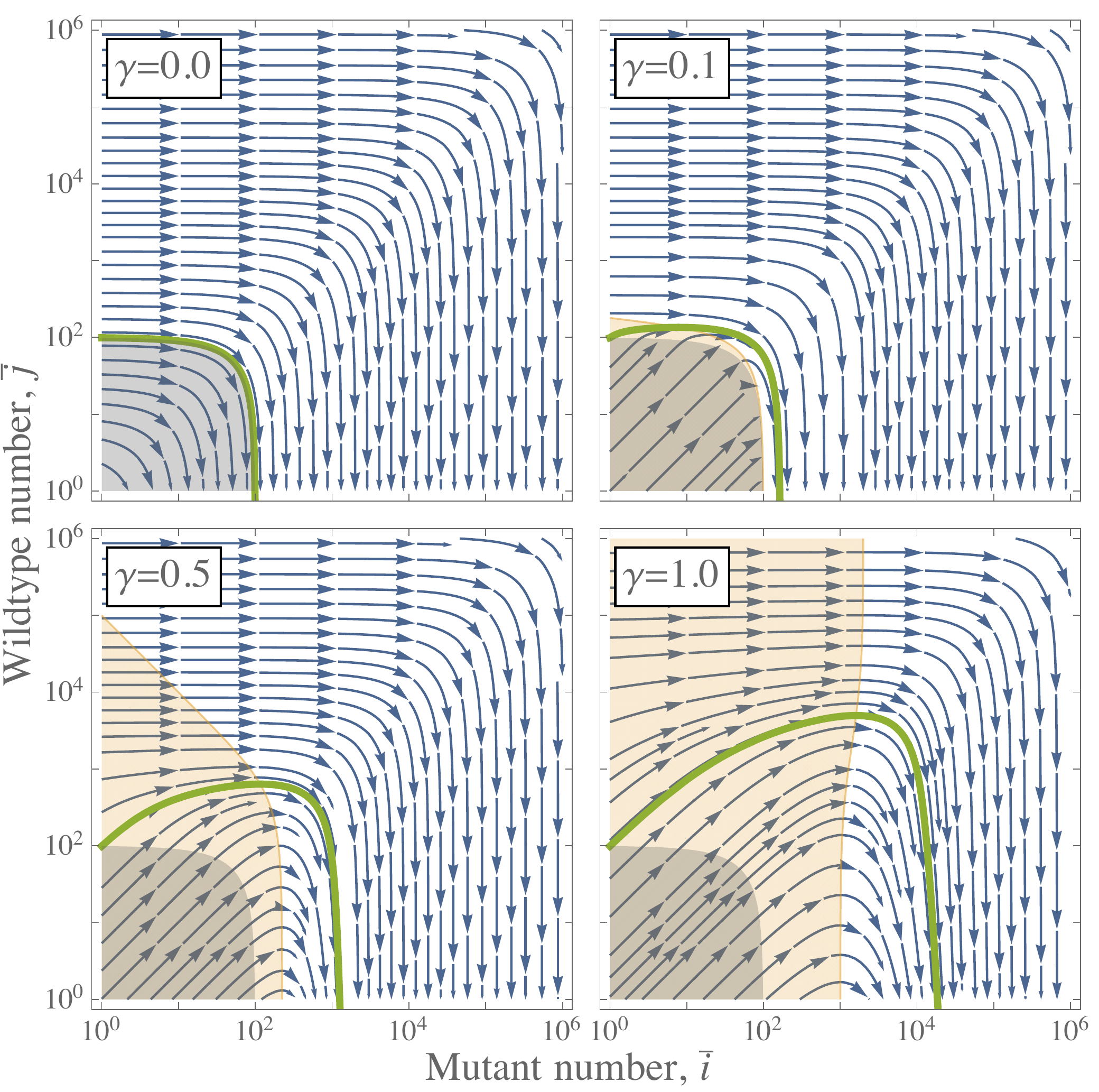}
\caption{
Deterministic flow for the dominance game in a growing population, as given by Eq.~\eqref{eq:detijPower}.
Payoff matrix is given by Eq.~\eqref{eq:payoffDominance}.
Here we consider no growth (top-left), weak (top-right), intermediate (bottom-left) and linear (bottom-right) growth.
Data shown is for $\beta = 0.1$ and $N_0=100$.
}
\label{fig:1-phaseDominance}
\end{figure}

\begin{figure}
\centering
\includegraphics[width=\columnwidth]{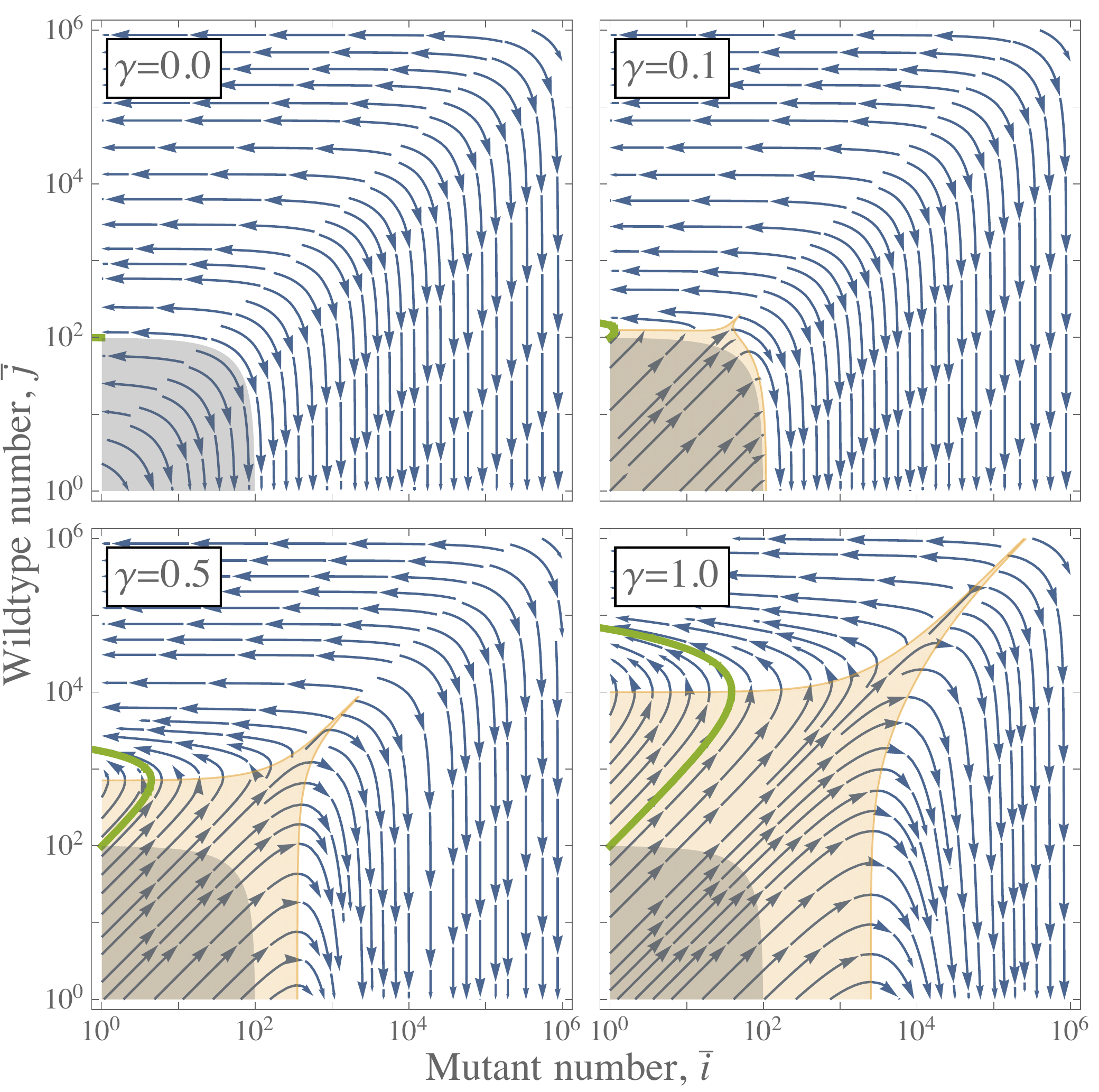}
\caption{
Deterministic flow for the coordination game in a growing population, as given by Eq.~\eqref{eq:detijPower}.
Payoff matrix is given by Eq.~\eqref{eq:payoffCoord} with $x^\star=0.2$.
Data shown is for $\beta = 0.1$ and $N_0=100$.
}
\label{fig:2-phaseCoord}
\end{figure}

\begin{figure}
\centering
\includegraphics[width=\columnwidth]{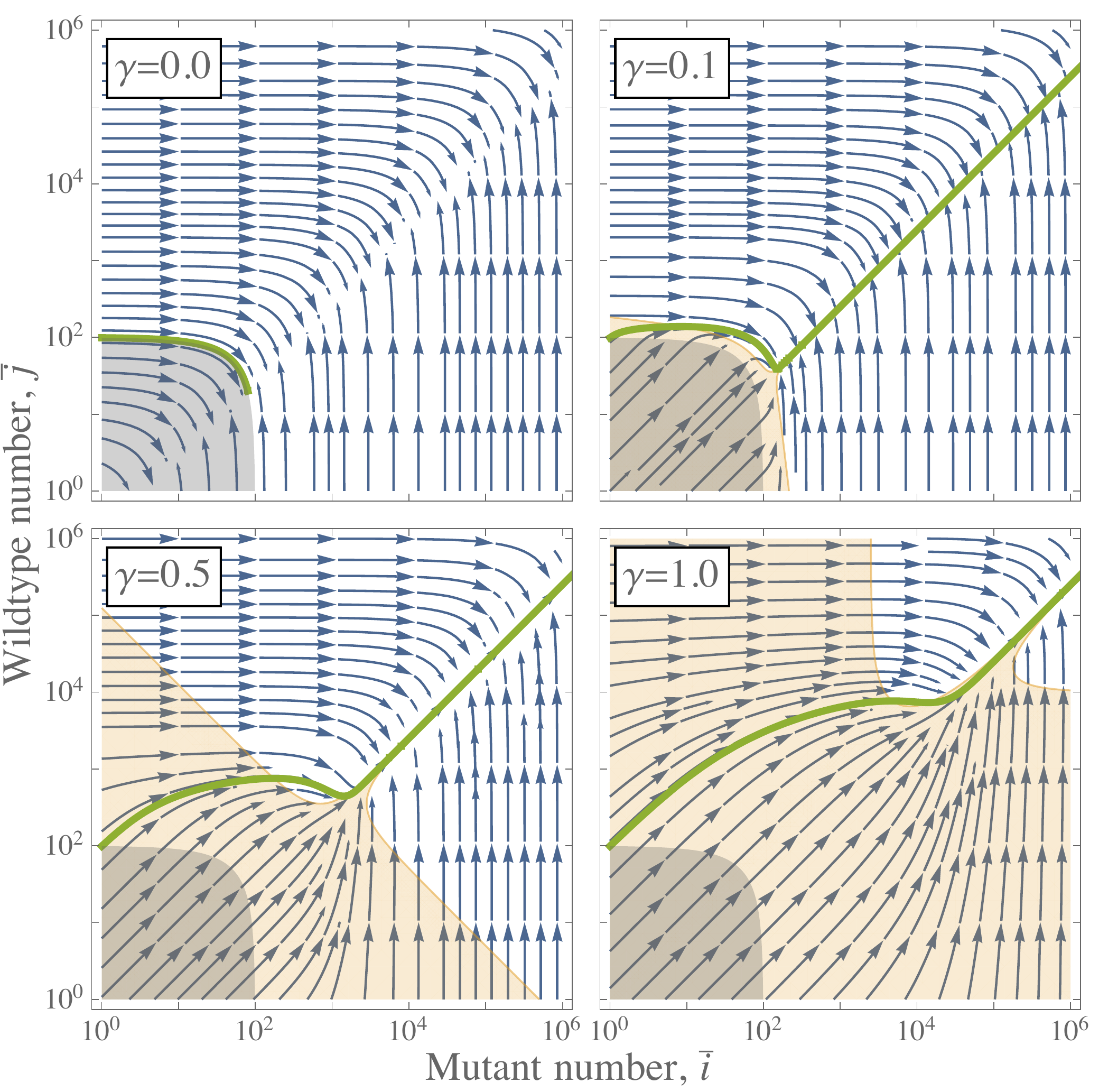}
\caption{
Deterministic flow for the coexistence game in a growing population, as given by Eq.~\eqref{eq:detijPower}.
Payoff matrix is as in Eq.~\eqref{eq:payoffCoex} with $x^\star = 0.8$.
Data shown is for $\beta = 0.1$ and $N_0=100$.
}
\label{fig:3-phaseCoex}
\end{figure}

In Figs.~\ref{fig:1-phaseDominance}, \ref{fig:2-phaseCoord}, and \ref{fig:3-phaseCoex} we show the deterministic flow field~\eqref{eq:detijPower} for dominance, coordination, and coexistence games for different growth exponents.
We describe each figure in turn below.
The dark-shaded regions in these figures correspond to $\olN < N_0$ (with $N_0=100$); these are situations which cannot be realised as the population starts at size $N_0$ and then increases.
It is, however, illustrative to show the flow in this region as well.
The light-shaded regions indicate the region where growth dominates over selection, i.e. $\dot{\oli}+\dot{\olj} > |\dot{\oli}-\dot{\olj}|$.
The thick, solid line indicates the deterministic trajectory of a single mutant in the wildtype population.

{\bf Dominance game:}
Here the fitness of one species is strictly higher than that of the other, irrespective of the composition of the population.
A payoff matrix which produces this dynamics is
\begin{equation} 
\begin{array}{c|c c} & A & B \\\hline A & 5 & 2 \\ B & 3 & 1 \end{array},
\label{eq:payoffDominance}
\end{equation}
for which one finds $\olpi_A(x) > \olpi_B(x)$ for all $x$.
Hence in this game (and in the absence of noise) the mutants will always eventually prevail ($\olj \to 0$).

The dominance of the mutants in this game is highlighted in Fig.~\ref{fig:1-phaseDominance}.
There is, however, an initial period in which both species grow in number.
This is a consequence of our choice of growth law which has an initial phase of rapid expansion, such that reproduction events dominate over selection in the early stages.
The initial expansion is greatest for large $\gamma$, and the region in which growth dominates also increases with $\gamma$.
When the population size is large, the dynamics are very similar to the no-growth case as selection dominates over growth for large $\olN$.

{\bf Coordination game:}
This scenario represents a case of bistability, and can be conveniently represented by the payoff matrix
\begin{equation}
\begin{array}{c|c c} & A & B \\\hline A & 1 & 1-x^\star \\ B & x^\star & 1 \end{array},
\label{eq:payoffCoord}
\end{equation}
where $0 < x^\star < 1$.
An individual's payoff is maximal if it interacts with another of its own type.
If mutant numbers are low, it is likely that they will have a low expected payoff and will be selected against.
However, if their numbers are large then they will have a higher payoff than the wildtype and will be selected for.
At some intermediate number there will be a `tipping point', where the direction of selection changes sign.
In the deterministic replicator dynamics~\eqref{eq:replicator} this unstable fixed point is located at $x^\star$.
The boundary states in which the mutant is extinct ($x=0$) or fixated ($x=1$) are both stable.

The resulting flow fields for the coordination game in growing populations is shown in Fig.~\ref{fig:2-phaseCoord}.
The two basins of attraction are separated by the unstable fixed point line (we use $x^\star = 0.2$ in the figure).
As in the dominance game the number of individuals of both types grow initially.
However, when the population size is large enough selection dominates over the growth, and the dynamics resembles the no-growth scenario.
This ultimately leads to the extinction of one species.
Under deterministic dynamics, the initial condition determines which one of the two species ultimately prevails; in this case the single mutant is destined to perish eventually.

{\bf Coexistence game:}
In a coexistence game the population is driven towards a heterogeneous state in which both species are present.
An individual's payoff is maximised if it interacts with the other species.
These games can be conveniently parameterised as
\begin{equation}
\begin{array}{c|c c} & A & B \\\hline A & 1 & 1+x^\star \\ B & 2-x^\star & 1,\end{array}
\label{eq:payoffCoex}
\end{equation}
where $0 < x^\star < 1$ is a stable fixed point under replicator dynamics~\eqref{eq:replicator}.
The boundary states in which the mutant is extinct ($x=0$) or fixated ($x=1$) are both unstable.
 
The corresponding flow fields are shown in Fig.~\ref{fig:3-phaseCoex} for different growth exponents and a fixed point of $x^\star = 0.8$.
As before both species grow provided their numbers are sufficiently low.
Selection does always act to maintain a heterogeneous population, however growth dominates for a large proportion of the $(\oli,\olj)$-plane (for large $\gamma$).
Under such deterministic dynamics no species would be lost from the system.
This outcome changes if intrinsic stochasticity is accounted for, however it becomes increasingly difficult to escape from the stable fixed point as the population size increases.


\section{Stochastic population dynamics} \label{sec:stoch}
We now turn to the outcome of the stochastic individual-based model.
We focus on the case in which one single mutant of type $A$ is initially present in the population of $N_0-1$ wildtypes.
Intrinsic noise in the birth--death dynamics then implies the possibility that this mutant may go extinct, or take over the entire population.
In populations of constant size one of these two outcomes will inevitably occur.
As we will see below, this is not always the case when the population grows.
Our analysis focuses on the fixation probability of the mutant, which we denote by $\phi_1$.
To obtain a systematic characterisation we investigate the different types of games separately.

\subsection{Dominance game}
In the dominance game described by the payoff matrix in Eq.~\eqref{eq:payoffDominance}, selection acts in favour of the mutant.
Growth of the population increases the strength of selection relative to the strength of the intrinsic noise, reducing the probability that the mutant goes extinct.
This in turn leads to a fixation probability which increases monotonically with the growth exponent $\gamma$.
We have verified this in simulations but do not show the data here, as it is relatively unspectacular.

Considering a mutant whose fitness is consistently lower than that of the wildtype we find that the probability of fixation monotonically decreases with the growth exponent.
We do not investigate these trivial results in more detail.

\subsection{Coordination game}
\begin{figure}
\centering
\includegraphics[width=\columnwidth]{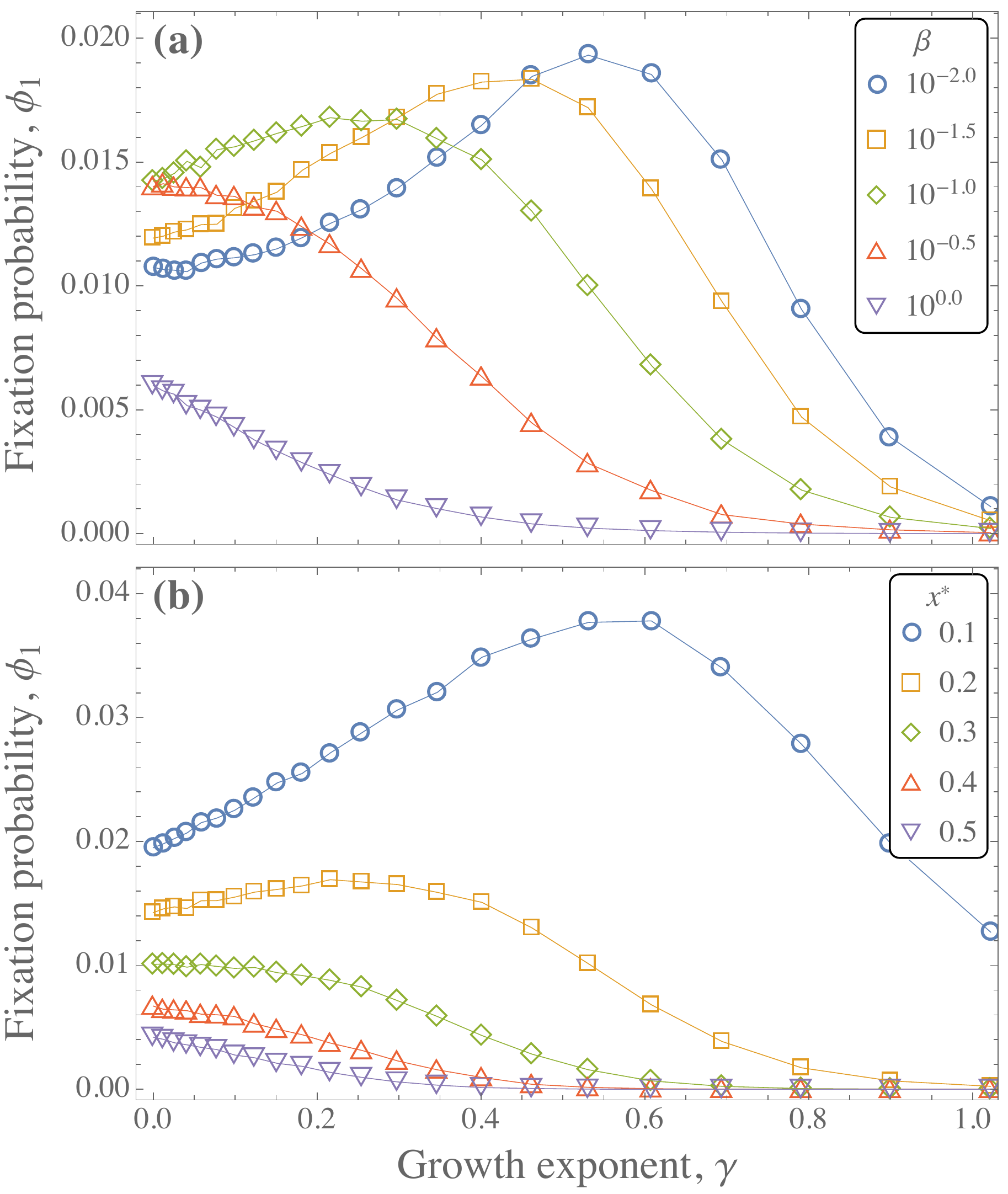}
\caption{
Probability for a single mutant to reach fixation in coordination games.
Symbols are data from $10^6$ simulations of the stochastic model, run until the mutant is extinct or has reached fixation.
Lines are guides to the eye.
{\bf (a)} The location of the fixed point is held fixed at $x^\star=0.2$, and the selection strength is varied between curves.
{\bf (b)} The selection strength is fixed at $\beta=0.1$, and different curves correspond to locations of the unstable fixed point of the replicator dynamics.
For all simulations the initial size of the population at $t=1$ is $N_0=100$; the payoff matrix is as in Eq.~\eqref{eq:payoffCoord}.
}
\label{fig:4-fixationCoord}
\end{figure}
We now turn to the scenario of coordination games, as defined by the payoff matrix in Eq.~\eqref{eq:payoffCoord}.
The fixation probability of a single mutant, $\phi_1$, in a population subject to power-law growth is shown in Fig.~\ref{fig:4-fixationCoord}.
We consider multiple combinations of the growth exponent $\gamma$, the selection strength $\beta$, and the location of the unstable fixed point $x^\star$.
The data in Fig.~\ref{fig:4-fixationCoord} shows that growth can have a non-trivial effect on the success of the mutant, provided that selection is sufficiently weak and that the fixed point is not too far from the extinction state.
The fixation probability of the mutant is then highest at intermediate speeds of growth.

An intuitive understanding of this behaviour can be obtained as follows: In the initial phases of the dynamics the population is small, and the effects of intrinsic noise dominate.
Different realisations of the stochastic process will lead to fractions of mutants spread across the interval $0<x<1$.
Crucially, these occupy both the region to the left of the unstable internal fixed point ($x<x^\star$), and the region to the right ($x>x^\star$).
With time the population grows, and so the underlying deterministic flow becomes stronger relative to the intrinsic noise.
Trajectories to the left of the fixed point will experience an increasing pull towards extinction of the mutant, whereas those to the right of $x^\star$ lead to its fixation.
 
For small growth exponents $\gamma$, the population size remains close to $N_0$ for a long time.
Thus the effects of selection do not set in in the early stages, and the mutant number can cross back and forth over the fixed point until one of the absorbing boundaries is reached.
The fixation probability is essentially that of a coordination game in a population of constant and relatively small size.

At moderate $\gamma$ the population undergoes an initial phase of relatively free exploration, populating the basins on both sides of the fixed point.
As the population grows the deterministic pull sets in.
Realisations that are to the right of the fixed point, and which could normally have diffused back and led to extinction of the mutant are prevented from doing so (statistically) and reach fixation instead ($x=1$).
Similarly realisations to the left of $x^\star$, which may ultimately have crossed the fixed point again and lead to fixation in the absence of growth, now lead to extinction of the mutant ($x=0$) due to the onset of deterministic pull.
If the fixed point is close to the extinction state the former effects outweighs the latter.
This results in an increased net probability for the mutant to reach fixation rather than extinction.

At large values of the growth exponent, the deterministic pull sets in very quickly.
The period of initial (nearly) free drift is short, and the mutants do not have time to expand and populate the basin to the right of the fixed point.
Broadly speaking the number of mutants remains small, and they quickly experience an increasing pull towards extinction.
The chances for the mutant to reach fixation are reduced.
In the extreme case of very quick growth ($\gamma \gtrsim 1$) the system becomes effectively deterministic immediately, hence $\phi_1 \approx 0$.
This is confirmed in Fig.~\ref{fig:4-fixationCoord}.

We next briefly discuss the effects of the selection strength, $\beta$.
As seen in Fig.~\ref{fig:4-fixationCoord}(a), increasing the selection strength moves the maximum in fixation probability towards lower values of the growth exponent $\gamma$.
This appears natural as increased selection indicates stronger deterministic flow at any given size of the population.
If selection is too strong the mutants are unlikely to overcome the initial barrier of adverse selection, even for moderate growth exponents.
We then find monotonically decreasing dependence of fixation probability on the growth exponent $\gamma$.

The above interpretation also suggests that the location of the internal fixed point might be relevant.
As described, the initial increase in fixation probability with the growth exponent is due to the  gradual `trapping' of realisations by the increasing deterministic pull. An increased fixation probability is thus only to be expected if the basin to the right of $x^\star$ is sufficiently large.
This is confirmed in Fig.~\ref{fig:4-fixationCoord}(b), where we demonstrate that the maximum in $\phi_1$ as a function of the growth exponent $\gamma$ is only present if $x^\star$ is sufficiently small.
If the fixed point is too far from the initial condition $x=1/N_0$ only very few runs are able to cross the barrier before the increasing deterministic pull confines them to the extinction basin.
Thus, when $x^\star$ is large, we find monotonically decreasing fixation probability as growth increases.

This interplay between $\phi_1$ and $x^\star$ is reminiscent of the so-called `one-third law' in evolutionary game theory, which is valid for populations of fixed finite size $N$ interacting in a coordination game.
The fixation probability of a single invading mutant in the limit of weak, but non-zero selection is then higher than that under neutral selection ($\phi=1/N$), if $x^\star<1/3$ \cite{nowak:nature:2004, ohtsuki:JTB:2007}.
We next investigate the impact of growth on this rule.

\begin{figure}
\centering
\includegraphics[width=\columnwidth]{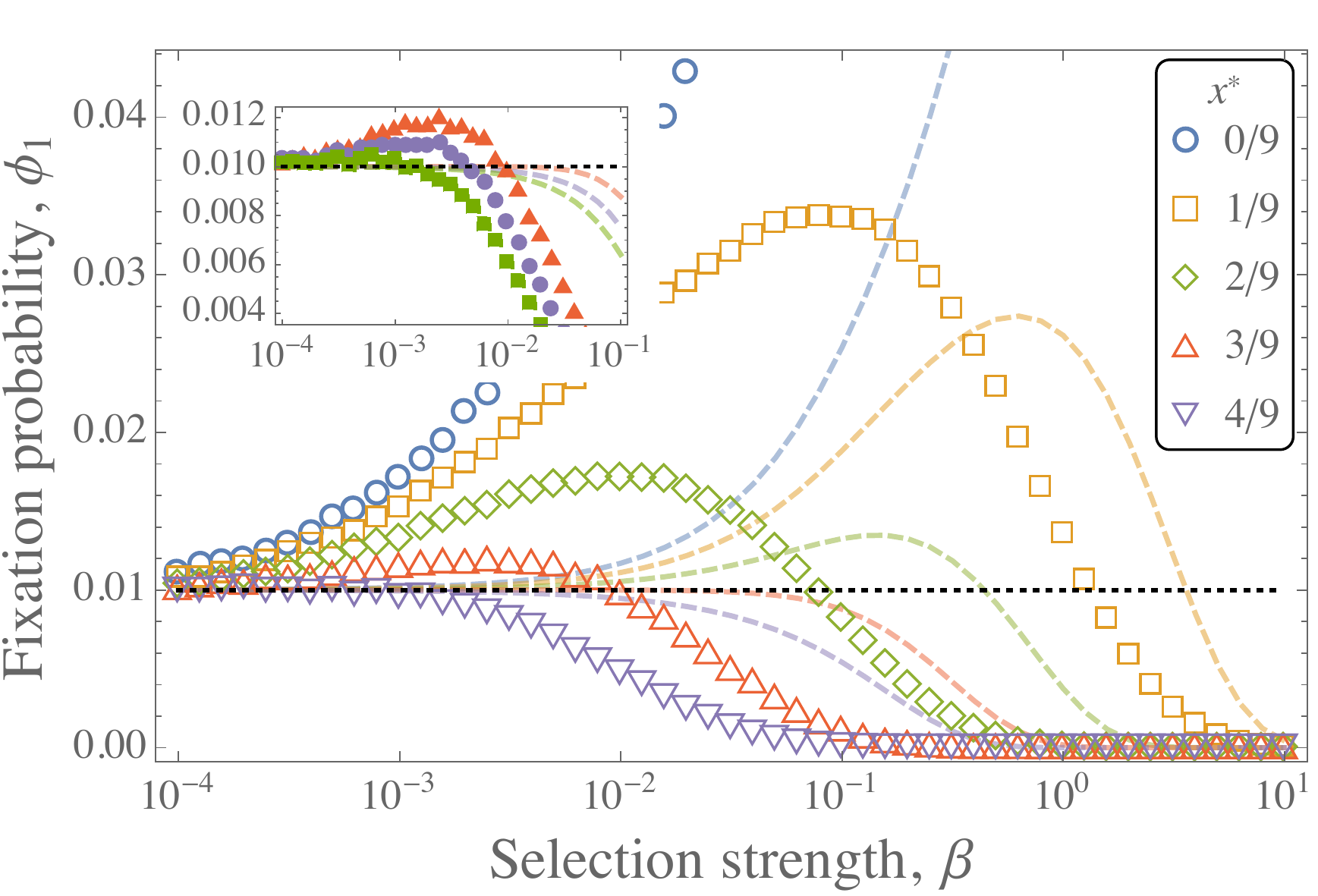}
\caption{
Breakdown of the one-third law of coordination games in a growing population.
Symbols show simulation results for the probability of a single mutant to reach fixation.
Data is obtained from $10^6$ simulations of the stochastic model, run until extinction or fixation of the mutant.
Dashed lines are the fixation probabilities in a population of fixed size $N=N_0$ ($x^\star=0,\,1/9,\,2/9,\,3/9,\,4/9$ from top to bottom), computed analytically by standard methods \cite{nowak:book:2006,traulsen:bookchapter:2009}.
Horizontal dotted line is the fixation probability $\phi_1=1/N_0$ under neutral selection in a population of constant size $N_0$.
In the inset we vary $x^\star$ more finely, with values $9/27$ (triangles), $10/27$ (circles), and $11/27$ (squares).
For all simulations the initial population at $t=1$ is $N_0=100$, the growth exponent is $\gamma=0.5$, and the payoff matrix is given by Eq.~\eqref{eq:payoffCoord}.
}
\label{fig:5-onethirdCoord}
\end{figure}

First, it is useful to re-formulate the one-third law for coordination games in a constant-size population in the following way: At small selection strengths $\beta$, the fixation probability $\phi_1$ is an increasing function of $\beta$ if $x^\star<1/3$, and it is a decreasing function otherwise.
Given that $\phi_1 \to 0$ in the limit $\beta\to\infty$ for coordination games, we expect a maximum in the function $\phi=\phi(\beta)$ when $x^\star<1/3$, and a monotonically decreasing function if $x^\star>1/3$.
This is verified by the dashed lines in Fig.~\ref{fig:5-onethirdCoord}, which show the analytical solution for the fixation probability in a population of fixed size $N=100$.
This solution is obtained by standard methods \cite{nowak:book:2006, traulsen:bookchapter:2009}. 
For $x^\star = 1/3$, the fixation probability is equal to the neutral result until $\beta$ is sufficiently large.

Simulations of a growing population reveal that the one-third law can break down when the size of the population is not fixed.
The symbols in Fig.~\ref{fig:5-onethirdCoord} are simulation data for $\phi_1$ in a population with growth exponent $\gamma=0.5$.
In the limit of weak-selection ($\beta\to 0$) we find $\phi_1 \to 1/N_0$, which is the result one obtains for neutral selection in a fixed-size population. The fixation probability of the mutant can increase with $\beta$, even when $x^\star \ge 1/3$.
This is highlighted in the inset of Fig.~\ref{fig:5-onethirdCoord}, where we vary $x^\star$ more finely.

\subsection{Coexistence game}
\begin{figure}
\centering
\includegraphics[width=\columnwidth]{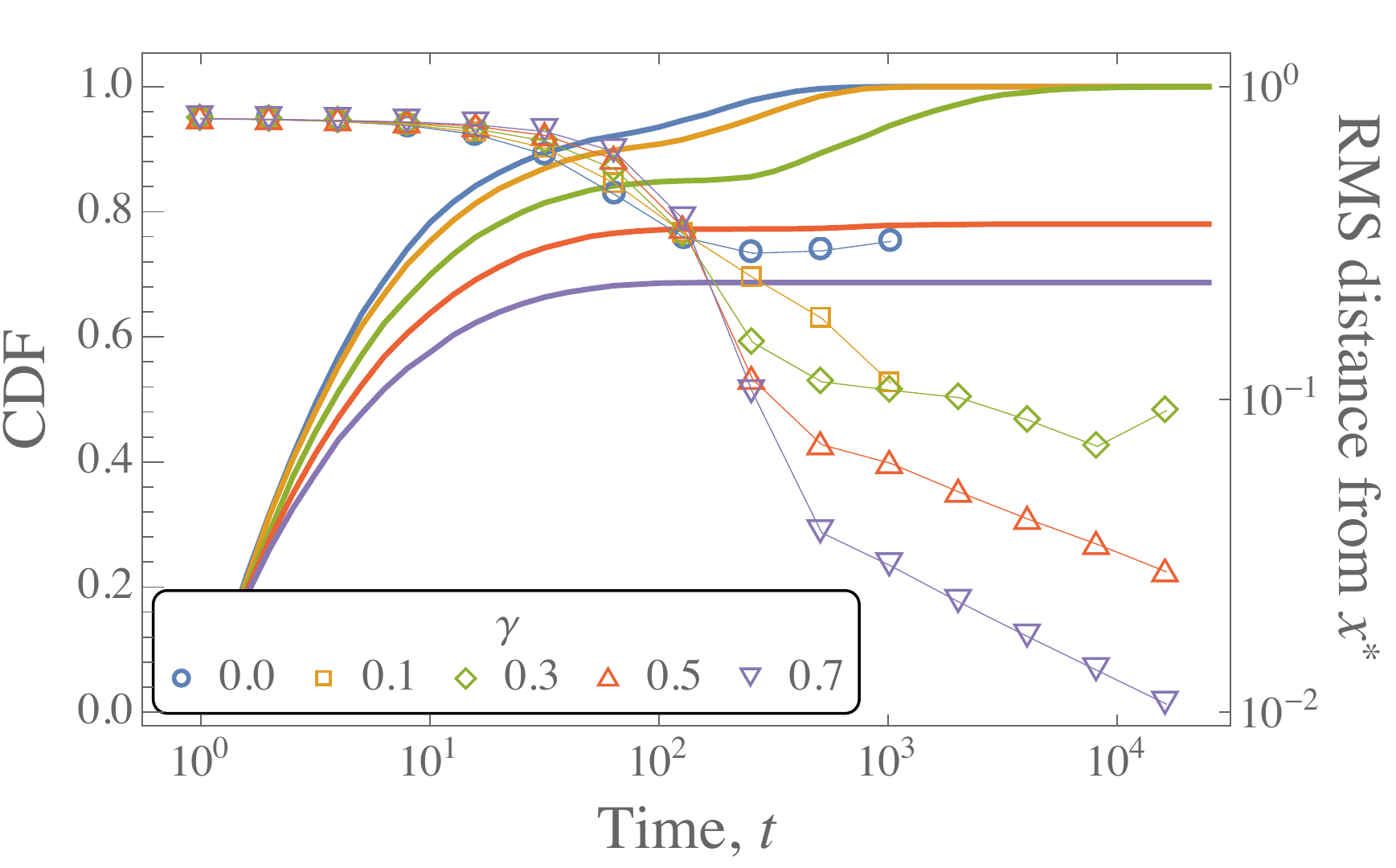}
\caption{
Stochastic dynamics of a single mutant in the coexistence game.
Solid lines show the probability to have reached either extinction or fixation by time $t$ (left axis; $\gamma=0$, $0.1$, $0.3$, $0.5$, $0.7$ from top to bottom).
Symbols show the root-mean-square of the distance from the stable fixed point $x^\star$ (right axis; see text for details).
Data is from $10^4$ simulations of the stochastic model, run until a maximum time $t=20,000$.
For all simulations the initial size of the population at $t=1$ is $N_0=100$, selection strength is $\beta=0.1$, the stable fixed point is located at $x^\star=0.8$, and the payoff matrix is as in Eq.~\eqref{eq:payoffCoex}.
}
\label{fig:6-cdfCoex}
\end{figure}
We now turn to the case of coexistence games in populations subject to power-law growth.
In this situation the deterministic replicator dynamics~\eqref{eq:replicator} has a stable internal fixed point, indicating species coexistence.
However, in finite populations of constant size an invading mutant will either go extinct or reach fixation due to the intrinsic stochasticity.
As in the case of the coordination game, the probability and mean time it takes to reach fixation or extinction in a constant-size population can be computed analytically from the backward master equation \cite{nowak:book:2006, traulsen:bookchapter:2009}.

We find that the situation changes when the population is allowed to grow.
With time the deterministic pull towards the internal stable state becomes stronger relative to random drift, and as a consequence fast-growing populations may never reach absorption.
This can be seen in Fig.~\ref{fig:0-trajectories}.

The simulation data shown as solid lines in Fig.~\ref{fig:6-cdfCoex} demonstrate this.
The lines indicate the fraction of simulation runs in which a single mutant has either reached fixation or gone extinct by time $t$, i.e. the cumulative distribution of absorption times (CDF).
This quantity reaches the value of one for $\gamma=0$ (constant population size), indicating that all runs reach either $x=0$ (extinction of the mutant) or $x=1$ (fixation) eventually.
The same is found for small, but positive values of the growth exponent $\gamma$ -- all samples reach an absorbing state eventually.
For faster-growing populations we find a finite fraction of samples in which the mutant neither reaches fixation nor goes extinct.
Our simulations naturally need to be stopped at some finite time, but as seen from the data in Fig.~\ref{fig:6-cdfCoex} the cumulative distribution of absorption times reaches a constant value (less than one) at finite times, with no further increase observed in the later parts of the simulations.
This confirms that the time horizon of our simulations is sufficiently long, and we can reasonably assume that no further extinction or fixation events would occur if the simulations were continued to later times.

To explain why coexistence can prevail indefinitely in growing populations, we focus on the root-mean-square distance from the stable fixed point $x^\star$. For each simulation run we record the fraction of mutants, $x(t) = i(t)/N(t)$ as a function of time, and then compute
\begin{equation}
{\rm RMS}(t) = \sqrt{ \avg{ \bigl[ x(t) - x^\star \bigr]^2 }_a},
\end{equation}
where $\avg{\cdot}_a$ represents the average over `active' simulation runs, i.e. runs that have not reached fixation or extinction before time $t$.
This quantity is a measure of how close, on average, the fraction of mutants is to the deterministic value $x^\star$.
It is plotted as symbols in Fig.~\ref{fig:6-cdfCoex}.
As can be seen, the RMS distance generally decreases with time; the population size is increasing such that the impact of intrinsic fluctuations decreases and deterministic effects dominate.
This confines the mutant fraction to the region around $x^\star$.
The escape probability is related to the width of the distribution (of $i/N$) about $x^\star$.
As this width decreases, so does the probability of escape.
The effect is stronger for faster-growing populations.

\begin{figure}
\centering
\includegraphics[width=\columnwidth]{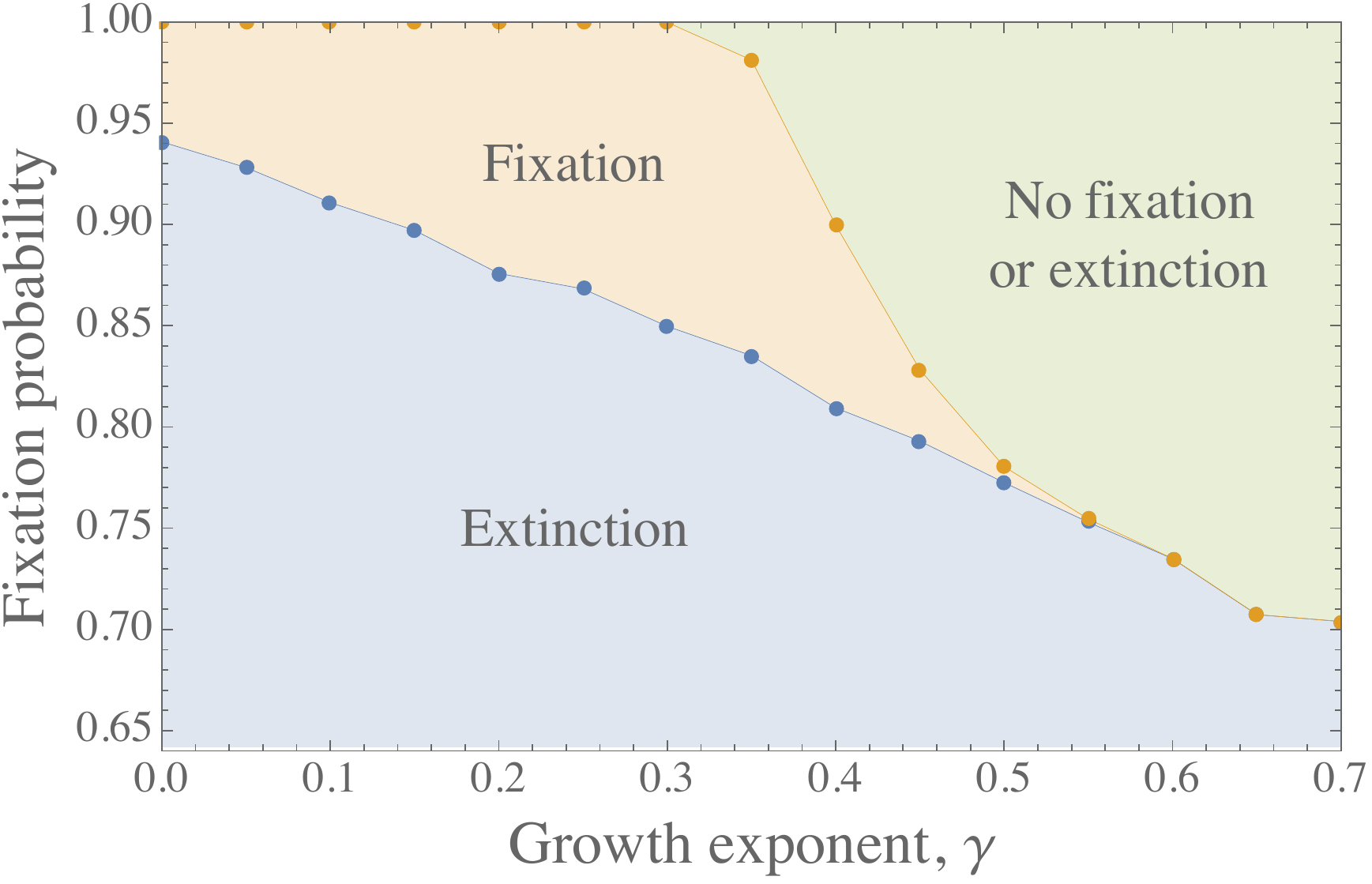}
\caption{
Stacked area chart for the probability of the different outcomes of a single mutant in the coexistence game.
The mutant may either reach fixation, go extinct, or the two species coexist until the end of the simulation.
Data is from $10^4$ simulations run until a maximum time of $t=20,000$.
The initial population at $t=1$ is $N_0=100$, selection strength is $\beta=0.1$, the stable fixed point is located at $x^\star=0.8$, and the payoff matrix is as in Eq.~\eqref{eq:payoffCoex}.
The vertical axis has been truncated.
}
\label{fig:7-fixationCoex}
\end{figure}

In the stacked area chart~\ref{fig:7-fixationCoex} we illustrate the outcome of introducing a single mutant into a wildtype population for different choices of the growth exponents $\gamma$.
For each realisation there are three possible outcomes: the initial mutant goes extinct, reaches fixation, or coexists with the wildtype until the end of the simulation (by extrapolation we then assume coexistence will continue indefinitely).
The figure reveals several characteristic features: the probability for the mutant to go extinct monotonically decreases as the growth exponent $\gamma$ is increased.
This is intuitively easy to understand, quicker growth implies that the deterministic pull of the coexistence game becomes relevant already in the initial stages of the simulation.
This drives the system towards the coexistence point, and away from extinction.
This also leads to an increased probability for the mutant to reach fixation at intermediate growth exponents, see Fig.~\ref{fig:7-fixationCoex}.
The system is driven to the coexistence point, but growth is not quick enough to eliminate the effects of random drift immediately.
Mutant numbers fluctuate around the coexistence point, and the population can then be driven to fixation at $x=1$ by intrinsic noise.

When $\gamma$ sufficiently large (fast growth), the latter step is inhibited.
The fraction of mutants will settle around $x^\star$ as described above, and the mutant type will be unable to overcome the selection barrier to reach fixation.
An increasing fraction of realisations is found to remain near the coexistence point indefinitely, and neither fixation nor extinction of the invading mutant takes place.

\begin{figure}
\centering
\includegraphics[width=\columnwidth]{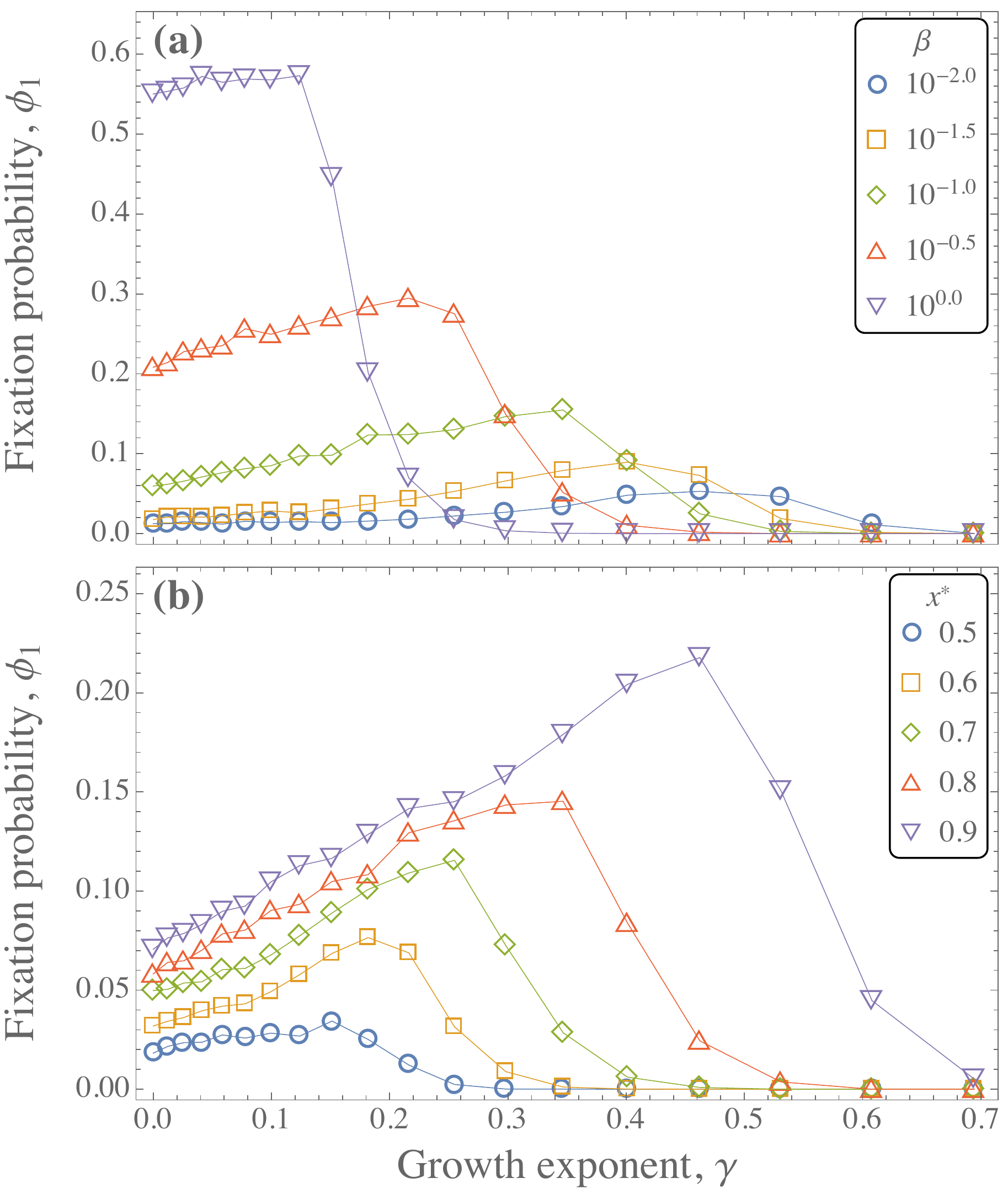}
\caption{
Probability for a single mutant to reach fixation in coexistence games.
Symbols are data from $10^4$ simulations of the stochastic model, run until time $t=20,000$.
Lines are guides to the eye.
{\bf (a)} The location of fixed point is held fixed at $x^\star = 0.8$, and the selection strength is varied between curves.
{\bf (b)} The selection strength is fixed at $\beta = 0.1$, and different curves correspond to locations of the stable fixed point.
The initial size of the population at $t=1$ is $N_0 = 100$; the payoff matrix is as in Eq.~\eqref{eq:payoffCoex}.
}
\label{fig:8-fixationCoexVary}
\end{figure}

Taken altogether these effects result in the behaviour of the fixation probability shown in Fig.~\ref{fig:8-fixationCoexVary}.
Strong selection helps the mutant avoid extinction, but combined with fast growth the deterministic pull towards $x^\star$ cannot be overcome and fixation cannot be reached.
With some growth the mutant has a higher chance of escaping extinction, and these combined effects lead to a maximum in the fixation probability $\phi_1$.
Provided the fixed point $x^\star$ is sufficiently close to the boundary at $x=1$ we find maximal fixation probability at intermediate growth rates $\gamma$.
This effect is stronger when the fixed point is close to $x=1$, and can be absent if the barrier between the coexistence point and fixation is too large (i.e. when $x^\star$ is too far to the left).


\section{Conclusions} \label{sec:concl}

In summary, we have investigated the effects of population growth on the outcome of stochastic evolutionary games.
While the stochastic dynamics of $2 \times 2$ evolutionary games in constant-size populations is largely understood, we find new features in growing populations.
To systematically study these effects we have disentangled growth and selection, and assume that both species in the population grow at the same (per capita) rate.
We impose an external time-dependence on these rates; while this can be interpreted as an experimental protocol in which the availability of nutrients is varied in time, our main objective is to generate specific growth laws of the population.
We focus on the case of power-law growth, $\olN(t)=N_0 t^\gamma$, and control the growth exponent $\gamma$.

The most interesting behaviour arises in coordination and coexistence games; we study scenarios in which a single mutant tries to invade a wildtype population.
We find that intermediate growth rates can lead to the maximum chance of success of the mutant; if growth is too fast or too slow (or non-existent), the ability for the mutant to invade is compromised.
This effect is present in both classes of games; the origin of this effect is different in the two cases, though.

In coordination games, the mutant will either become extinct or reach fixation eventually.
Population growth increases the effect of selection over time relative to intrinsic noise.
In the later stages of the process this prevents the mutant number from crossing the barrier separating the basins of attraction of extinction and fixation, and shifts the balance between the two outcomes.
These effects also lead to the breakdown of the one-third law of coordination games.

In coexistence games, a growing population can promote indefinite coexistence of the mutant and wildtype individuals.
By effectively decreasing the magnitude of intrinsic fluctuations, population growth confines the fraction of mutants to the region near the deterministic coexistence point.
If the growth rate is moderate, selection is strong enough initially to prevent extinction but leaves the population size small enough to be able to eventually escape from coexistence to fixation of the mutant.

In the appendix we propose an approximation of the dynamics in terms of a single stochastic differential equation.
Here the amplitude of multiplicative noise decreases gradually in time, representing the increasing population size.
This equation is formulated ad-hoc and we make no claims of analytical rigour.
However, we find good quantitative agreement with direct simulations.
This framework allows efficient computation when population sizes are large, and may lead to further analytical progress.

Introducing a variable population size has revealed interesting effects in many evolutionary models \cite{melbinger:PRL:2010,constable:PNAS:2016,chotibut:PRE:2015}.
We have continued this line of work and systematically investigated the fate of mutants in a growing population.
In biological settings populations grow -- this is often their primary function.
Adding growth to well-established evolutionary models thus helps to bridge the gap between experimental work on growing and evolving populations, and theoretical understanding.
One next step might be developing models in which the growth-law emerges dynamically, for example through environmental processes acting on the population.
If such an emergent law matches our externally-imposed growth dynamics, then we expect our results to hold, at least qualitatively.
 
On a general level, our work contributes to the increasing literature that tries to understand the interplay between deterministic selection and intrinsic stochasticity.
Varying the intensity of noise by introducing population growth is a novel way to approach this task.

\begin{acknowledgments}
TG would like to thank the Group of Nonlinear Physics, University of Santiago de Compostela, Spain for hospitality. 
\end{acknowledgments}


\begin{thebibliography}{36}%
\makeatletter
\providecommand \@ifxundefined [1]{%
 \@ifx{#1\undefined}
}%
\providecommand \@ifnum [1]{%
 \ifnum #1\expandafter \@firstoftwo
 \else \expandafter \@secondoftwo
 \fi
}%
\providecommand \@ifx [1]{%
 \ifx #1\expandafter \@firstoftwo
 \else \expandafter \@secondoftwo
 \fi
}%
\providecommand \natexlab [1]{#1}%
\providecommand \enquote  [1]{``#1''}%
\providecommand \bibnamefont  [1]{#1}%
\providecommand \bibfnamefont [1]{#1}%
\providecommand \citenamefont [1]{#1}%
\providecommand \href@noop [0]{\@secondoftwo}%
\providecommand \href [0]{\begingroup \@sanitize@url \@href}%
\providecommand \@href[1]{\@@startlink{#1}\@@href}%
\providecommand \@@href[1]{\endgroup#1\@@endlink}%
\providecommand \@sanitize@url [0]{\catcode `\\12\catcode `\$12\catcode
  `\&12\catcode `\#12\catcode `\^12\catcode `\_12\catcode `\%12\relax}%
\providecommand \@@startlink[1]{}%
\providecommand \@@endlink[0]{}%
\providecommand \url  [0]{\begingroup\@sanitize@url \@url }%
\providecommand \@url [1]{\endgroup\@href {#1}{\urlprefix }}%
\providecommand \urlprefix  [0]{URL }%
\providecommand \Eprint [0]{\href }%
\providecommand \doibase [0]{http://dx.doi.org/}%
\providecommand \selectlanguage [0]{\@gobble}%
\providecommand \bibinfo  [0]{\@secondoftwo}%
\providecommand \bibfield  [0]{\@secondoftwo}%
\providecommand \translation [1]{[#1]}%
\providecommand \BibitemOpen [0]{}%
\providecommand \bibitemStop [0]{}%
\providecommand \bibitemNoStop [0]{.\EOS\space}%
\providecommand \EOS [0]{\spacefactor3000\relax}%
\providecommand \BibitemShut  [1]{\csname bibitem#1\endcsname}%
\let\auto@bib@innerbib\@empty
\bibitem [{\citenamefont {Maddamsetti}\ \emph {et~al.}(2015)\citenamefont
  {Maddamsetti}, \citenamefont {Lenski},\ and\ \citenamefont
  {Barrick}}]{maddamsetti:Genetics:2015}%
  \BibitemOpen
  \bibfield  {author} {\bibinfo {author} {\bibfnamefont {R.}~\bibnamefont
  {Maddamsetti}}, \bibinfo {author} {\bibfnamefont {R.~E.}\ \bibnamefont
  {Lenski}}, \ and\ \bibinfo {author} {\bibfnamefont {J.~E.}\ \bibnamefont
  {Barrick}},\ }\href@noop {} {\bibfield  {journal} {\bibinfo  {journal}
  {Genetics}\ }\textbf {\bibinfo {volume} {200}},\ \bibinfo {pages} {619}
  (\bibinfo {year} {2015})}\BibitemShut {NoStop}%
\bibitem [{\citenamefont {Nowak}(2006)}]{nowak:book:2006}%
  \BibitemOpen
  \bibfield  {author} {\bibinfo {author} {\bibfnamefont {M.~A.}\ \bibnamefont
  {Nowak}},\ }\href@noop {} {\emph {\bibinfo {title} {Evolutionary dynamics}}}\
  (\bibinfo  {publisher} {Harvard University Press, Cambridge MA},\ \bibinfo
  {year} {2006})\BibitemShut {NoStop}%
\bibitem [{\citenamefont {Altrock}\ \emph {et~al.}(2015)\citenamefont
  {Altrock}, \citenamefont {Liu},\ and\ \citenamefont
  {Michor}}]{altrock:NatRev:2015}%
  \BibitemOpen
  \bibfield  {author} {\bibinfo {author} {\bibfnamefont {P.~M.}\ \bibnamefont
  {Altrock}}, \bibinfo {author} {\bibfnamefont {L.~L.}\ \bibnamefont {Liu}}, \
  and\ \bibinfo {author} {\bibfnamefont {F.}~\bibnamefont {Michor}},\
  }\href@noop {} {\bibfield  {journal} {\bibinfo  {journal} {Nat. Rev. Cancer}\
  }\textbf {\bibinfo {volume} {15}},\ \bibinfo {pages} {730} (\bibinfo {year}
  {2015})}\BibitemShut {NoStop}%
\bibitem [{\citenamefont {Castellano}\ \emph {et~al.}(2009)\citenamefont
  {Castellano}, \citenamefont {Fortunato},\ and\ \citenamefont
  {Loreto}}]{castellano:RMP:2009}%
  \BibitemOpen
  \bibfield  {author} {\bibinfo {author} {\bibfnamefont {C.}~\bibnamefont
  {Castellano}}, \bibinfo {author} {\bibfnamefont {S.}~\bibnamefont
  {Fortunato}}, \ and\ \bibinfo {author} {\bibfnamefont {V.}~\bibnamefont
  {Loreto}},\ }\href@noop {} {\bibfield  {journal} {\bibinfo  {journal} {Rev.
  Mod. Phys.}\ }\textbf {\bibinfo {volume} {81}},\ \bibinfo {pages} {591}
  (\bibinfo {year} {2009})}\BibitemShut {NoStop}%
\bibitem [{\citenamefont {Hofbauer}\ and\ \citenamefont
  {Sigmund}(1998)}]{hofbauer:book:1998}%
  \BibitemOpen
  \bibfield  {author} {\bibinfo {author} {\bibfnamefont {J.}~\bibnamefont
  {Hofbauer}}\ and\ \bibinfo {author} {\bibfnamefont {K.}~\bibnamefont
  {Sigmund}},\ }\href@noop {} {\emph {\bibinfo {title} {Evolutionary Games and
  Population Dynamics}}}\ (\bibinfo  {publisher} {Cambridge University Press,
  Cambridge UK},\ \bibinfo {year} {1998})\BibitemShut {NoStop}%
\bibitem [{\citenamefont {Maynard~Smith}(1982)}]{maynard-smith:book:1982}%
  \BibitemOpen
  \bibfield  {author} {\bibinfo {author} {\bibfnamefont {J.}~\bibnamefont
  {Maynard~Smith}},\ }\href@noop {} {\emph {\bibinfo {title} {Evolution and the
  Theory of Games}}}\ (\bibinfo  {publisher} {Cambridge University Press,
  Cambridge UK},\ \bibinfo {year} {1982})\BibitemShut {NoStop}%
\bibitem [{\citenamefont {Traulsen}\ and\ \citenamefont
  {Hauert}(2009)}]{traulsen:bookchapter:2009}%
  \BibitemOpen
  \bibfield  {author} {\bibinfo {author} {\bibfnamefont {A.}~\bibnamefont
  {Traulsen}}\ and\ \bibinfo {author} {\bibfnamefont {C.}~\bibnamefont
  {Hauert}},\ }in\ \href@noop {} {\emph {\bibinfo {booktitle} {Reviews of
  Nonlinear Dynamics and Complexity}}},\ Vol.~\bibinfo {volume} {II},\ \bibinfo
  {editor} {edited by\ \bibinfo {editor} {\bibfnamefont {H.~G.}\ \bibnamefont
  {Schuster}}}\ (\bibinfo  {publisher} {Wiley-VCH, Weinheim},\ \bibinfo {year}
  {2009})\ pp.\ \bibinfo {pages} {25--61}\BibitemShut {NoStop}%
\bibitem [{\citenamefont {Ewens}(2004)}]{ewens:book:2004}%
  \BibitemOpen
  \bibfield  {author} {\bibinfo {author} {\bibfnamefont {W.~J.}\ \bibnamefont
  {Ewens}},\ }\href@noop {} {\emph {\bibinfo {title} {Mathematical Population
  Genetics 1: Theoretical Introduction}}}\ (\bibinfo  {publisher} {Springer,
  New York},\ \bibinfo {year} {2004})\BibitemShut {NoStop}%
\bibitem [{\citenamefont {Goel}\ and\ \citenamefont
  {Richter-Dyn}(1974)}]{goel:book:1974}%
  \BibitemOpen
  \bibfield  {author} {\bibinfo {author} {\bibfnamefont {N.}~\bibnamefont
  {Goel}}\ and\ \bibinfo {author} {\bibfnamefont {N.}~\bibnamefont
  {Richter-Dyn}},\ }\href@noop {} {\emph {\bibinfo {title} {Stochastic Models
  in Biology}}}\ (\bibinfo  {publisher} {Academic Press, New York},\ \bibinfo
  {year} {1974})\BibitemShut {NoStop}%
\bibitem [{\citenamefont {Taylor}\ \emph {et~al.}(2004)\citenamefont {Taylor},
  \citenamefont {Fudenberg}, \citenamefont {Sasaki},\ and\ \citenamefont
  {Nowak}}]{taylor:BMB:2004}%
  \BibitemOpen
  \bibfield  {author} {\bibinfo {author} {\bibfnamefont {C.}~\bibnamefont
  {Taylor}}, \bibinfo {author} {\bibfnamefont {D.}~\bibnamefont {Fudenberg}},
  \bibinfo {author} {\bibfnamefont {A.}~\bibnamefont {Sasaki}}, \ and\ \bibinfo
  {author} {\bibfnamefont {M.~A.}\ \bibnamefont {Nowak}},\ }\href@noop {}
  {\bibfield  {journal} {\bibinfo  {journal} {Bull. Math. Biol.}\ }\textbf
  {\bibinfo {volume} {66}},\ \bibinfo {pages} {1621} (\bibinfo {year}
  {2004})}\BibitemShut {NoStop}%
\bibitem [{\citenamefont {Bladon}\ \emph {et~al.}(2010)\citenamefont {Bladon},
  \citenamefont {Galla},\ and\ \citenamefont {McKane}}]{bladon:PRE:2010}%
  \BibitemOpen
  \bibfield  {author} {\bibinfo {author} {\bibfnamefont {A.~J.}\ \bibnamefont
  {Bladon}}, \bibinfo {author} {\bibfnamefont {T.}~\bibnamefont {Galla}}, \
  and\ \bibinfo {author} {\bibfnamefont {A.~J.}\ \bibnamefont {McKane}},\
  }\href@noop {} {\bibfield  {journal} {\bibinfo  {journal} {Phys. Rev. E}\
  }\textbf {\bibinfo {volume} {81}},\ \bibinfo {pages} {066122} (\bibinfo
  {year} {2010})}\BibitemShut {NoStop}%
\bibitem [{\citenamefont {Gardiner}(2009)}]{gardiner:book:2009}%
  \BibitemOpen
  \bibfield  {author} {\bibinfo {author} {\bibfnamefont {C.~W.}\ \bibnamefont
  {Gardiner}},\ }\href@noop {} {\emph {\bibinfo {title} {Handbook of Stochastic
  Methods}}}\ (\bibinfo  {publisher} {Springer, New York},\ \bibinfo {year}
  {2009})\BibitemShut {NoStop}%
\bibitem [{\citenamefont {van Kampen}(2007)}]{kampen:book:2007}%
  \BibitemOpen
  \bibfield  {author} {\bibinfo {author} {\bibfnamefont {N.~G.}\ \bibnamefont
  {van Kampen}},\ }\href@noop {} {\emph {\bibinfo {title} {Stochastic Processes
  in Physics and Chemistry}}}\ (\bibinfo  {publisher} {Elsevier, Amsterdam},\
  \bibinfo {year} {2007})\BibitemShut {NoStop}%
\bibitem [{\citenamefont {Altrock}\ and\ \citenamefont
  {Traulsen}(2009)}]{altrock:NJP:2009}%
  \BibitemOpen
  \bibfield  {author} {\bibinfo {author} {\bibfnamefont {P.~M.}\ \bibnamefont
  {Altrock}}\ and\ \bibinfo {author} {\bibfnamefont {A.}~\bibnamefont
  {Traulsen}},\ }\href@noop {} {\bibfield  {journal} {\bibinfo  {journal} {New
  J. Phys.}\ }\textbf {\bibinfo {volume} {11}},\ \bibinfo {pages} {013012}
  (\bibinfo {year} {2009})}\BibitemShut {NoStop}%
\bibitem [{\citenamefont {Antal}\ and\ \citenamefont
  {Scheuring}(2006)}]{antal:BMB:2006}%
  \BibitemOpen
  \bibfield  {author} {\bibinfo {author} {\bibfnamefont {T.}~\bibnamefont
  {Antal}}\ and\ \bibinfo {author} {\bibfnamefont {I.}~\bibnamefont
  {Scheuring}},\ }\href@noop {} {\bibfield  {journal} {\bibinfo  {journal}
  {Bull. Math. Biol.}\ }\textbf {\bibinfo {volume} {68}},\ \bibinfo {pages}
  {1923} (\bibinfo {year} {2006})}\BibitemShut {NoStop}%
\bibitem [{\citenamefont {Melbinger}\ \emph {et~al.}(2010)\citenamefont
  {Melbinger}, \citenamefont {Cremer},\ and\ \citenamefont
  {Frey}}]{melbinger:PRL:2010}%
  \BibitemOpen
  \bibfield  {author} {\bibinfo {author} {\bibfnamefont {A.}~\bibnamefont
  {Melbinger}}, \bibinfo {author} {\bibfnamefont {J.}~\bibnamefont {Cremer}}, \
  and\ \bibinfo {author} {\bibfnamefont {E.}~\bibnamefont {Frey}},\ }\href@noop
  {} {\bibfield  {journal} {\bibinfo  {journal} {Phys. Rev. Lett.}\ }\textbf
  {\bibinfo {volume} {105}},\ \bibinfo {pages} {178101} (\bibinfo {year}
  {2010})}\BibitemShut {NoStop}%
\bibitem [{\citenamefont {Cremer}\ \emph {et~al.}(2011)\citenamefont {Cremer},
  \citenamefont {Melbinger},\ and\ \citenamefont {Frey}}]{cremer:PRE:2011}%
  \BibitemOpen
  \bibfield  {author} {\bibinfo {author} {\bibfnamefont {J.}~\bibnamefont
  {Cremer}}, \bibinfo {author} {\bibfnamefont {A.}~\bibnamefont {Melbinger}}, \
  and\ \bibinfo {author} {\bibfnamefont {E.}~\bibnamefont {Frey}},\ }\href@noop
  {} {\bibfield  {journal} {\bibinfo  {journal} {Phys. Rev. E}\ }\textbf
  {\bibinfo {volume} {84}},\ \bibinfo {pages} {051921} (\bibinfo {year}
  {2011})}\BibitemShut {NoStop}%
\bibitem [{\citenamefont {Cremer}\ \emph {et~al.}(2012)\citenamefont {Cremer},
  \citenamefont {Melbinger},\ and\ \citenamefont {Frey}}]{cremer:SciRep:2012}%
  \BibitemOpen
  \bibfield  {author} {\bibinfo {author} {\bibfnamefont {J.}~\bibnamefont
  {Cremer}}, \bibinfo {author} {\bibfnamefont {A.}~\bibnamefont {Melbinger}}, \
  and\ \bibinfo {author} {\bibfnamefont {E.}~\bibnamefont {Frey}},\ }\href@noop
  {} {\bibfield  {journal} {\bibinfo  {journal} {Sci. Rep}\ }\textbf {\bibinfo
  {volume} {2}},\ \bibinfo {pages} {281} (\bibinfo {year} {2012})}\BibitemShut
  {NoStop}%
\bibitem [{\citenamefont {Novak}\ \emph {et~al.}(2013)\citenamefont {Novak},
  \citenamefont {Chatterjee},\ and\ \citenamefont {Nowak}}]{novak:JTB:2013}%
  \BibitemOpen
  \bibfield  {author} {\bibinfo {author} {\bibfnamefont {S.}~\bibnamefont
  {Novak}}, \bibinfo {author} {\bibfnamefont {K.}~\bibnamefont {Chatterjee}}, \
  and\ \bibinfo {author} {\bibfnamefont {M.~A.}\ \bibnamefont {Nowak}},\
  }\href@noop {} {\bibfield  {journal} {\bibinfo  {journal} {J. Theor. Biol.}\
  }\textbf {\bibinfo {volume} {334}},\ \bibinfo {pages} {26} (\bibinfo {year}
  {2013})}\BibitemShut {NoStop}%
\bibitem [{\citenamefont {Chotibut}\ and\ \citenamefont
  {Nelson}(2015)}]{chotibut:PRE:2015}%
  \BibitemOpen
  \bibfield  {author} {\bibinfo {author} {\bibfnamefont {T.}~\bibnamefont
  {Chotibut}}\ and\ \bibinfo {author} {\bibfnamefont {D.~R.}\ \bibnamefont
  {Nelson}},\ }\href@noop {} {\bibfield  {journal} {\bibinfo  {journal} {Phys.
  Rev. E}\ }\textbf {\bibinfo {volume} {92}},\ \bibinfo {pages} {022718}
  (\bibinfo {year} {2015})}\BibitemShut {NoStop}%
\bibitem [{\citenamefont {Huang}\ \emph {et~al.}(2015)\citenamefont {Huang},
  \citenamefont {Hauert},\ and\ \citenamefont {Traulsen}}]{huang:PNAS:2015}%
  \BibitemOpen
  \bibfield  {author} {\bibinfo {author} {\bibfnamefont {W.}~\bibnamefont
  {Huang}}, \bibinfo {author} {\bibfnamefont {C.}~\bibnamefont {Hauert}}, \
  and\ \bibinfo {author} {\bibfnamefont {A.}~\bibnamefont {Traulsen}},\
  }\href@noop {} {\bibfield  {journal} {\bibinfo  {journal} {Proc. Natl. Acad.
  Sci. U.S.A.}\ }\textbf {\bibinfo {volume} {112}},\ \bibinfo {pages} {9064}
  (\bibinfo {year} {2015})}\BibitemShut {NoStop}%
\bibitem [{\citenamefont {Li}\ \emph {et~al.}(2015)\citenamefont {Li},
  \citenamefont {Pietschke}, \citenamefont {Fraune}, \citenamefont {Altrock},
  \citenamefont {Bosch},\ and\ \citenamefont {Traulsen}}]{li:Interface:2015}%
  \BibitemOpen
  \bibfield  {author} {\bibinfo {author} {\bibfnamefont {X.-Y.}\ \bibnamefont
  {Li}}, \bibinfo {author} {\bibfnamefont {C.}~\bibnamefont {Pietschke}},
  \bibinfo {author} {\bibfnamefont {S.}~\bibnamefont {Fraune}}, \bibinfo
  {author} {\bibfnamefont {P.~M.}\ \bibnamefont {Altrock}}, \bibinfo {author}
  {\bibfnamefont {T.~C.~G.}\ \bibnamefont {Bosch}}, \ and\ \bibinfo {author}
  {\bibfnamefont {A.}~\bibnamefont {Traulsen}},\ }\href@noop {} {\bibfield
  {journal} {\bibinfo  {journal} {J. R. Soc. Interface}\ }\textbf {\bibinfo
  {volume} {12}},\ \bibinfo {pages} {20150121} (\bibinfo {year}
  {2015})}\BibitemShut {NoStop}%
\bibitem [{\citenamefont {Constable}\ \emph {et~al.}(2016)\citenamefont
  {Constable}, \citenamefont {Rogers}, \citenamefont {McKane},\ and\
  \citenamefont {Tarnita}}]{constable:PNAS:2016}%
  \BibitemOpen
  \bibfield  {author} {\bibinfo {author} {\bibfnamefont {G.~W.}\ \bibnamefont
  {Constable}}, \bibinfo {author} {\bibfnamefont {T.}~\bibnamefont {Rogers}},
  \bibinfo {author} {\bibfnamefont {A.~J.}\ \bibnamefont {McKane}}, \ and\
  \bibinfo {author} {\bibfnamefont {C.~E.}\ \bibnamefont {Tarnita}},\
  }\href@noop {} {\bibfield  {journal} {\bibinfo  {journal} {Proc. Natl. Acad.
  Sci. U.S.A.}\ }\textbf {\bibinfo {volume} {113}},\ \bibinfo {pages}
  {201603693} (\bibinfo {year} {2016})}\BibitemShut {NoStop}%
\bibitem [{\citenamefont {Papkou}\ \emph {et~al.}(2016)\citenamefont {Papkou},
  \citenamefont {Gokhale}, \citenamefont {Traulsen},\ and\ \citenamefont
  {Schulenburg}}]{papkou:Zoo:2016}%
  \BibitemOpen
  \bibfield  {author} {\bibinfo {author} {\bibfnamefont {A.}~\bibnamefont
  {Papkou}}, \bibinfo {author} {\bibfnamefont {C.~S.}\ \bibnamefont {Gokhale}},
  \bibinfo {author} {\bibfnamefont {A.}~\bibnamefont {Traulsen}}, \ and\
  \bibinfo {author} {\bibfnamefont {H.}~\bibnamefont {Schulenburg}},\
  }\href@noop {} {\bibfield  {journal} {\bibinfo  {journal} {Zoology}\ }\textbf
  {\bibinfo {volume} {119}},\ \bibinfo {pages} {330} (\bibinfo {year}
  {2016})}\BibitemShut {NoStop}%
\bibitem [{\citenamefont {Morris}\ and\ \citenamefont
  {Rogers}(2014)}]{morris:JPhysA:2014}%
  \BibitemOpen
  \bibfield  {author} {\bibinfo {author} {\bibfnamefont {R.~G.}\ \bibnamefont
  {Morris}}\ and\ \bibinfo {author} {\bibfnamefont {T.}~\bibnamefont
  {Rogers}},\ }\href@noop {} {\bibfield  {journal} {\bibinfo  {journal} {J.
  Phys. A}\ }\textbf {\bibinfo {volume} {47}},\ \bibinfo {pages} {342003}
  (\bibinfo {year} {2014})}\BibitemShut {NoStop}%
\bibitem [{\citenamefont {Hallatschek}\ \emph {et~al.}(2007)\citenamefont
  {Hallatschek}, \citenamefont {Hersen}, \citenamefont {Ramanathan},\ and\
  \citenamefont {Nelson}}]{hallatschek:PNAS:2007}%
  \BibitemOpen
  \bibfield  {author} {\bibinfo {author} {\bibfnamefont {O.}~\bibnamefont
  {Hallatschek}}, \bibinfo {author} {\bibfnamefont {P.}~\bibnamefont {Hersen}},
  \bibinfo {author} {\bibfnamefont {S.}~\bibnamefont {Ramanathan}}, \ and\
  \bibinfo {author} {\bibfnamefont {D.~R.}\ \bibnamefont {Nelson}},\
  }\href@noop {} {\bibfield  {journal} {\bibinfo  {journal} {Proc. Natl. Acad.
  Sci. U.S.A.}\ }\textbf {\bibinfo {volume} {104}},\ \bibinfo {pages} {19926}
  (\bibinfo {year} {2007})}\BibitemShut {NoStop}%
\bibitem [{\citenamefont {Cappuccio}\ \emph {et~al.}(2006)\citenamefont
  {Cappuccio}, \citenamefont {Elishmereni},\ and\ \citenamefont
  {Agur}}]{cappuccio:CancerRes:2006}%
  \BibitemOpen
  \bibfield  {author} {\bibinfo {author} {\bibfnamefont {A.}~\bibnamefont
  {Cappuccio}}, \bibinfo {author} {\bibfnamefont {M.}~\bibnamefont
  {Elishmereni}}, \ and\ \bibinfo {author} {\bibfnamefont {Z.}~\bibnamefont
  {Agur}},\ }\href@noop {} {\bibfield  {journal} {\bibinfo  {journal} {Cancer
  Res.}\ }\textbf {\bibinfo {volume} {66}},\ \bibinfo {pages} {7293} (\bibinfo
  {year} {2006})}\BibitemShut {NoStop}%
\bibitem [{\citenamefont {Br\'u}\ \emph {et~al.}(2003)\citenamefont {Br\'u},
  \citenamefont {Albertos}, \citenamefont {Subiza}, \citenamefont
  {Garc\'ia-Asenjo},\ and\ \citenamefont {Br\'u}}]{bru:BPJ:2003}%
  \BibitemOpen
  \bibfield  {author} {\bibinfo {author} {\bibfnamefont {A.}~\bibnamefont
  {Br\'u}}, \bibinfo {author} {\bibfnamefont {S.}~\bibnamefont {Albertos}},
  \bibinfo {author} {\bibfnamefont {J.~L.}\ \bibnamefont {Subiza}}, \bibinfo
  {author} {\bibfnamefont {J.~L.}\ \bibnamefont {Garc\'ia-Asenjo}}, \ and\
  \bibinfo {author} {\bibfnamefont {I.}~\bibnamefont {Br\'u}},\ }\href@noop {}
  {\bibfield  {journal} {\bibinfo  {journal} {Biophys. J.}\ }\textbf {\bibinfo
  {volume} {85}},\ \bibinfo {pages} {2948} (\bibinfo {year}
  {2003})}\BibitemShut {NoStop}%
\bibitem [{\citenamefont {Karev}(2014)}]{karev:MathBiosci:2014}%
  \BibitemOpen
  \bibfield  {author} {\bibinfo {author} {\bibfnamefont {G.~P.}\ \bibnamefont
  {Karev}},\ }\href@noop {} {\bibfield  {journal} {\bibinfo  {journal} {Math.
  Biosci.}\ }\textbf {\bibinfo {volume} {258}},\ \bibinfo {pages} {85}
  (\bibinfo {year} {2014})}\BibitemShut {NoStop}%
\bibitem [{\citenamefont {Szathm{\'a}ry}\ and\ \citenamefont {{Maynard
  Smith}}(1997)}]{szathmary:JTB:1997}%
  \BibitemOpen
  \bibfield  {author} {\bibinfo {author} {\bibfnamefont {E.}~\bibnamefont
  {Szathm{\'a}ry}}\ and\ \bibinfo {author} {\bibfnamefont {J.}~\bibnamefont
  {{Maynard Smith}}},\ }\href@noop {} {\bibfield  {journal} {\bibinfo
  {journal} {J. Theor. Biol.}\ }\textbf {\bibinfo {volume} {187}},\ \bibinfo
  {pages} {555} (\bibinfo {year} {1997})}\BibitemShut {NoStop}%
\bibitem [{\citenamefont {Redner}(2001)}]{redner:book:2001}%
  \BibitemOpen
  \bibfield  {author} {\bibinfo {author} {\bibfnamefont {S.}~\bibnamefont
  {Redner}},\ }\href@noop {} {\emph {\bibinfo {title} {A guide to first-passage
  processes}}}\ (\bibinfo  {publisher} {Cambridge University Press, Camdridge
  UK},\ \bibinfo {year} {2001})\BibitemShut {NoStop}%
\bibitem [{\citenamefont {Nowak}\ \emph {et~al.}(2004)\citenamefont {Nowak},
  \citenamefont {Sasaki}, \citenamefont {Taylor},\ and\ \citenamefont
  {Fudenberg}}]{nowak:nature:2004}%
  \BibitemOpen
  \bibfield  {author} {\bibinfo {author} {\bibfnamefont {M.~A.}\ \bibnamefont
  {Nowak}}, \bibinfo {author} {\bibfnamefont {A.}~\bibnamefont {Sasaki}},
  \bibinfo {author} {\bibfnamefont {C.}~\bibnamefont {Taylor}}, \ and\ \bibinfo
  {author} {\bibfnamefont {D.}~\bibnamefont {Fudenberg}},\ }\href@noop {}
  {\bibfield  {journal} {\bibinfo  {journal} {Nature}\ }\textbf {\bibinfo
  {volume} {428}},\ \bibinfo {pages} {646} (\bibinfo {year}
  {2004})}\BibitemShut {NoStop}%
\bibitem [{\citenamefont {Ohtsuki}\ \emph {et~al.}(2007)\citenamefont
  {Ohtsuki}, \citenamefont {Bordalo},\ and\ \citenamefont
  {Nowak}}]{ohtsuki:JTB:2007}%
  \BibitemOpen
  \bibfield  {author} {\bibinfo {author} {\bibfnamefont {H.}~\bibnamefont
  {Ohtsuki}}, \bibinfo {author} {\bibfnamefont {P.}~\bibnamefont {Bordalo}}, \
  and\ \bibinfo {author} {\bibfnamefont {M.~A.}\ \bibnamefont {Nowak}},\
  }\href@noop {} {\bibfield  {journal} {\bibinfo  {journal} {J. Theor. Biol.}\
  }\textbf {\bibinfo {volume} {249}},\ \bibinfo {pages} {289} (\bibinfo {year}
  {2007})}\BibitemShut {NoStop}%
\bibitem [{\citenamefont {Traulsen}\ \emph {et~al.}(2012)\citenamefont
  {Traulsen}, \citenamefont {Claussen},\ and\ \citenamefont
  {Hauert}}]{traulsen:PRE:2012}%
  \BibitemOpen
  \bibfield  {author} {\bibinfo {author} {\bibfnamefont {A.}~\bibnamefont
  {Traulsen}}, \bibinfo {author} {\bibfnamefont {J.~C.}\ \bibnamefont
  {Claussen}}, \ and\ \bibinfo {author} {\bibfnamefont {C.}~\bibnamefont
  {Hauert}},\ }\href@noop {} {\bibfield  {journal} {\bibinfo  {journal} {Phys.
  Rev. E}\ }\textbf {\bibinfo {volume} {85}},\ \bibinfo {pages} {041901}
  (\bibinfo {year} {2012})}\BibitemShut {NoStop}%
\bibitem [{\citenamefont {Traulsen}\ \emph {et~al.}(2005)\citenamefont
  {Traulsen}, \citenamefont {Claussen},\ and\ \citenamefont
  {Hauert}}]{traulsen:PRL:2005}%
  \BibitemOpen
  \bibfield  {author} {\bibinfo {author} {\bibfnamefont {A.}~\bibnamefont
  {Traulsen}}, \bibinfo {author} {\bibfnamefont {J.~C.}\ \bibnamefont
  {Claussen}}, \ and\ \bibinfo {author} {\bibfnamefont {C.}~\bibnamefont
  {Hauert}},\ }\href@noop {} {\bibfield  {journal} {\bibinfo  {journal} {Phys.
  Rev. Lett.}\ }\textbf {\bibinfo {volume} {95}},\ \bibinfo {pages} {238701}
  (\bibinfo {year} {2005})}\BibitemShut {NoStop}%
\bibitem [{\citenamefont {Hanson}(2007)}]{hanson:book:2007}%
  \BibitemOpen
  \bibfield  {author} {\bibinfo {author} {\bibfnamefont {F.~B.}\ \bibnamefont
  {Hanson}},\ }\href@noop {} {\emph {\bibinfo {title} {Applied stochastic
  processes and control for Jump-diffusions: modeling, analysis, and
  computation}}}\ (\bibinfo  {publisher} {SIAM},\ \bibinfo {year}
  {2007})\BibitemShut {NoStop}%
\end{thebibliography}
%

\appendix*

\section{Approximation by stochastic differential equation} \label{sec:SDE}

\begin{figure*}[bp]
\centering
\includegraphics[width=0.95\textwidth]{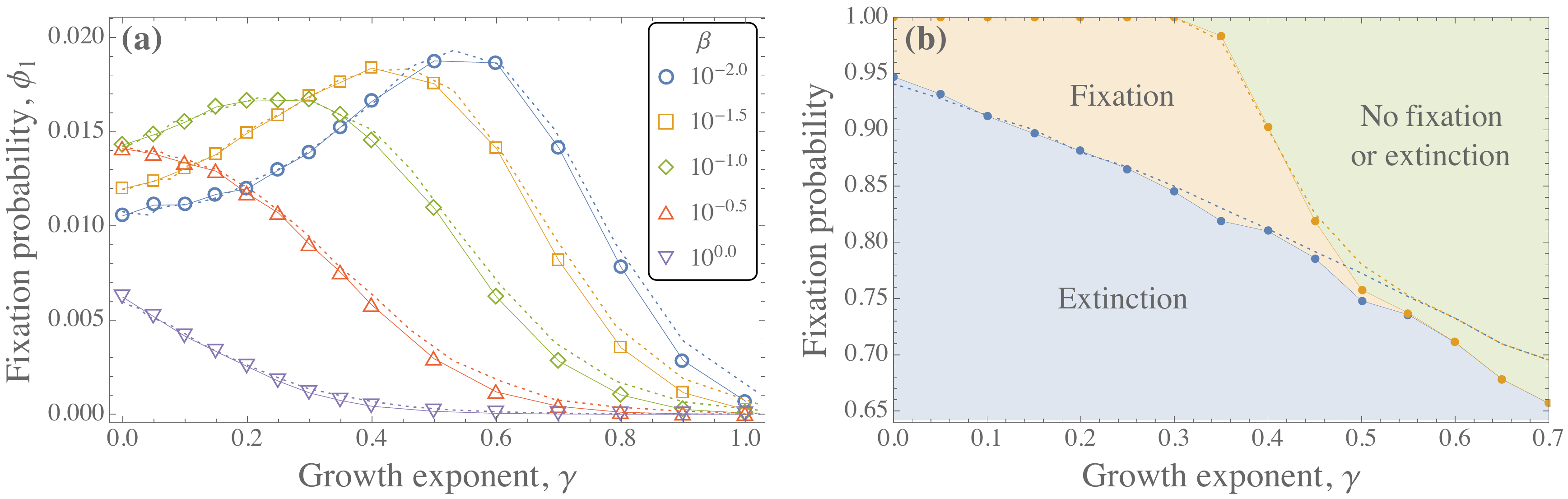} 
\caption{
Results from numerical integration of the SDE~\eqref{eq:sde}.
Simulations are started at $x=0.01$ at time $t=1$ to represent the case in which a single mutant is placed in a population of initial size $N_0=100$.
{\bf (a)} Probability for the mutant to fixate in coordination games.
This is the analog of Fig.~\ref{fig:4-fixationCoord}(a).
{\bf (b)} Probabilities of the different outcomes at the end of a simulation in coexistence games.
This is the analog of Fig.~\ref{fig:7-fixationCoex}.
Timestep of the numerical integration is $\Delta t = 0.001$.
Dotted lines are analogous results from the individual-based model.
}
\label{fig:9-sde}
\end{figure*}

The dynamics of large but finite populations can often be described by stochastic differential equations (SDE).
These resulting SDEs can formally be derived using a Kramers--Moyal or system-size expansion, and a well-defined formalism is available to do this \cite{kampen:book:2007,gardiner:book:2009}.
For non-constant populations this SDE approach has been used, for example, in \cite{melbinger:PRL:2010,cremer:PRE:2011,cremer:SciRep:2012,chotibut:PRE:2015}.
Such a procedure can have numerous benefits: the approximation of an individual-based process by an SDE can lead to significant computational speed-ups \cite{traulsen:PRE:2012}, which is especially important in our scenario where the growing population can quickly reach very large numbers.

In our model there are two degrees of freedom, which we write as $x(t)=i(t)/N(t)$ and $N(t)$.
Our aim here is not to rigorously derive SDEs for $x$ and $N$ (such a derivation is beyond the scope of this article), but to test the viability of a simpler phenomenological approach which has a significantly lower computational cost.
Specifically, we simulate the process
\begin{linenomath}
\begin{align}
\dot{x} &=  x(1-x)\left[g(\olpi_A,\olpi_B) - g(\olpi_B,\olpi_A)\right] \nonumber\\
&\quad + \sqrt{\frac{ x(1-x)\left[g(\olpi_A,\olpi_B) + g(\olpi_B,\olpi_A)\right] }{ \olN(t) }}\eta(t),
\label{eq:sde}
\end{align}
\end{linenomath}
where $\olN(t) = N_0 t^\gamma$ is the deterministic system size, and where $\eta(t)$ is Gaussian white noise of unit amplitude.
This equation is motivated by the well-known outcome for the diffusion approximation in populations of constant size $N$, as derived for example in \cite{traulsen:PRL:2005}.
Ignoring the noise term, we recover the replicator equation \eqref{eq:replicator}.

In Fig.~\ref{fig:9-sde} we show data from a simple numerical integration of Eq.~\eqref{eq:sde} (Euler--Maruyama scheme with constant timestep).
In our approach we terminate simulations once the variable $x$ leaves the interval $(0,1)$, and identify these realisations as fixated ($x \ge 1$) or extinct ($x \le 0$).
Despite the ad-hoc nature of Eq.~\eqref{eq:sde}, the data in Fig.~\ref{fig:9-sde} demonstrates that this approach is sufficient to capture the main features: Simulations of the coordination game [Fig.~\ref{fig:9-sde}(a); to be compared with Fig.~\ref{fig:4-fixationCoord}(a)] and the coexistence game [Fig.~\ref{fig:9-sde}(b); to be compared with Fig.~\ref{fig:7-fixationCoex}] show good agreement with individual-based simulations, even at the quantitative level.
We attribute small quantitative deviations to the approximations made in formulating the SDE, to the discretisation used in integrating it, and to artifacts in the numerical treatment of the multiplicative noise and absorbing boundaries.

The natural next step would be to attempt to analytically determine the fixation probabilities directly from the SDE~\eqref{eq:sde}.
While standard techniques are available for autonomous SDEs (i.e., those without external time dependence) \cite{gardiner:book:2009,hanson:book:2007}, the problem is more intricate in our case due to the explicit time-dependence of the noise amplitude.
While a backward Fokker--Planck equation can still be formulated (and solved numerically), an analytical characterisation of fixation times remains an open challenge.

\end{document}